\documentclass[aps,pre,superscriptaddress,tightenlines,onecolumn,showpacs,preprint,
floatfix]{revtex4-1}
\usepackage{graphicx}
\usepackage{verbatim}
\usepackage{graphics}
\usepackage{longtable} 
\usepackage{subfigure}
\usepackage{amsmath}
\usepackage{wrapfig}
\usepackage{epsfig}
\usepackage{float}
\usepackage{graphicx}
\usepackage{array}
\usepackage{psfrag}
\usepackage{color}
\usepackage{bbold}
\usepackage{tikz}
\usepackage[colorlinks, linkcolor=blue,citecolor=blue, urlcolor=blue]{hyperref}

\usepackage{color}

\begin{document}
\title{Liquid phase separation controlled by pH}

\author{Omar Adame-Arana}
\affiliation{Max-Planck-Institut f\"ur Physik komplexer Systeme, N\"othnitzer Str. 38, 01187 Dresden, 
Germany}

\author{Christoph A. Weber}
\affiliation{Max-Planck-Institut f\"ur Physik komplexer Systeme, N\"othnitzer Str. 38, 01187 Dresden, 
Germany}
\affiliation{Center for Systems Biology Dresden, Pfotenhauerstr. 108, 01307 Dresden, Germany}

\author{Vasily Zaburdaev}
\affiliation{Friedrich-Alexander Universit\"at Erlangen-N\"urnberg, Cauerstr. 11, 91058 Erlangen, 
Germany}
\affiliation{Max-Planck-Zentrum f\"ur Physik und Medizin, Staudtstr. 2, 91058 Erlangen, Germany}

\author{Jacques Prost}
\affiliation{Laboratoire Physico Chimie Curie, Institut Curie, PSL Research University, CNRS UMR168, 
75005 Paris, France}
\affiliation{Mechanobiology Institute, National University of Singapore, 5A Engineering 
Drive 1, Singapore 117411, Singapore}

\author{Frank J\"ulicher}
\affiliation{Max-Planck-Institut f\"ur Physik komplexer Systeme, N\"othnitzer Str. 38, 01187 Dresden, 
Germany}
\affiliation{Center for Systems Biology Dresden, Pfotenhauerstr. 108, 01307 Dresden, Germany}
\affiliation{Cluster of Excellence Physics of Life, Technische Universit\"at Dresden, 01062 Dresden, 
Germany}
\date{\today}
\begin{abstract}
We present a minimal model to study liquid phase separation in a fixed pH ensemble. The model describes 
a mixture composed of macromolecules that exist in three different charge states and have a tendency to 
phase separate. We introduce the pH dependence of phase separation by means of a set of reactions 
describing the protonation and deprotonation of macromolecules, as well as the self-ionisation of 
water. We use conservation laws to identify the conjugate thermodynamic variables at chemical 
equilibrium. Using this thermodynamic conjugate variables we perform a Legendre transform 
which defines the corresponding free energy at fixed pH. We first study the possible phase diagram 
topologies at the isoelectric point of the macromolecules. We then show how the phase behavior depends 
on pH by moving away from the isoelectric point. We find that phase diagrams as a function of pH 
strongly depend on whether oppositely charged macromolecules or neutral macromolecules have a stronger 
tendency to phase separate. We predict the existence of reentrant behavior as a function of pH. In 
addition, our model also predicts that the region of phase separation is typically broader at the 
isoelectric point. This model could account for both, the protein separation observed in yeast cells for 
pH values close to the isoelectric point of many cytosolic proteins and also for the {\it{in vitro}} 
experiments of single proteins exhibiting phase separation as a function of pH. 
\end{abstract}

\maketitle

\section{Introduction}
One of the central challenges in biology is to understand the spatial organisation of cells. Cells 
are organised in distinct compartments which provide specific biochemical environments 
that play a role for many physico-chemical processes such as the production of ATP or the assembly 
of cellular building blocks~\cite{alberts2017molecular}. In many cases the cell achieves spatial 
organisation of its biochemistry by forming membrane-less organelles, such as P granules, stress 
granules and centrosomes \cite{brangwynne2009germline,molliex2015phase,Zwicker2014}. It has been 
shown in recent years that many membrane-less organelles have liquid like properties. They are 
condensates of certain types of protein and often RNA that can be best understood as forming by a 
phase separation process~\cite{brangwynne2009germline, brangwynne2013phase, elbaum2015disordered, 
zhu2015nuclear,banani2017biomolecular}. 

While some of these compartments persist for longer times, others, such as stress granules form 
rapidly in response to stimulus such as  changes in temperature, pH or depletion of 
nutrients~\cite{cellular_fitness_Alberti_2018}. The formation of liquid-like condensates can also 
be reconstituted {\it{in vitro}} using purified proteins\cite{Patel2015,saha2016polar}.

The formation of protein condensates is also important in other contexts. The cytoplasm of yeast 
cells was shown to transition from a fluid-like to an arrested solid-like state during nutrient 
depletion~\cite{liquid_to_solid_Alberti_2016}. This transition is reversible and provides a 
protective mechanism that helps cells to survive periods of nutrient depletion until conditions 
improve and nutrients become available again. The biophysical mechanism responsible for this 
transition has been linked to changes in the cytosolic pH. The effects of pH on yeast cells can be 
described as follows. During nutrient depletion, a yeast cell does not have sufficient resources to 
supply proton pumps that are responsible for regulating the intracellular pH. As a result, in an 
acidic environment, natural for yeast habitat, the pH of the cytoplasm drops and many cytosolic proteins 
become insoluble. If the resulting protein condensates occupy a large volume fraction of the cytoplasm,  
the cytoplasm can transition from a fluid-like to a solid-like arrested 
state. This observation suggests that the reduction of pH triggers phase separation of proteins from 
solution. Interestingly, in this case, phase separation is triggered as the pH of the solution moves 
closer to the isoelectric points of many cytosolic proteins. This raises the question why pH changes, 
and in particular, pH values in the vicinity of the isoelectric point, promote phase separation. More 
generally we want to understand how the formation of protein condensates can be regulated by changes in 
pH.

To address this question we present a generic thermodynamic framework to study the influence of pH 
on liquid-liquid phase separation. The key idea is to couple a system capable of undergoing phase 
separation with a set of chemical reactions corresponding to the protonation/deprotonation of water 
components and macromolecules such as proteins. We consider two types of interactions, namely 
attractive interactions between oppositely charged macromolecules in the presence of counterions 
and salt, and attractive interactions among neutral macromolecules that could be mediated for 
example by Van der Waals or hydrophobic interactions. Using conservation laws and chemical equilibrium 
conditions~\cite{alberty1994legendre}, we construct an effective thermodynamic potential describing a 
system with pH as a thermodynamic variable. We then use this thermodynamic potential to determine the 
phase behaviour of the system as a function of the molecular properties and pH. We find coexisting 
phases of different compositions of charged and uncharged macromolecules. We show that the compositions 
of the coexisting phases can be controlled by changing pH.

The manuscript is organised as follows. In Section~\ref{sect:influence_of_pH}, we introduce a set 
of chemical reactions in which the charge state of a macromolecule is fixed by the pH of the 
system. We then define the pH and show its relation to the previously introduced chemical reactions. In
Section~\ref{sect:thermodynamics_of_multicomponent_mixtures} we present the thermodynamics of 
multicomponent mixtures and discuss the parameter choices for our study. We study the thermodynamic 
equilibrium for a system with fixed pH in Section~\ref{sect:chem_eq_at_fixed_pH}, whereby using 
conservation laws we identify the thermodynamic conjugate variables of the system which we then use 
to construct the corresponding thermodynamic potential for fixed pH. In 
Section~\ref{sect:control_of_phase_separation_by_pH}, we introduce new composition variables and 
thermodynamic fields controlling phase separation and discuss chemical and phase equilibrium in 
terms of the newly defined variables. In Sections~\ref{sect:phase_diagrams_at_pI} and 
\ref{sect:phase_coexistence_away_from_pH}, we study the phase behaviour at and 
away from the isoelectric point respectively. Finally, we discuss our results in 
Section~\ref{sect:discussion}.

\section{Chemical reactions and pH in macromolecular systems}\label{sect:influence_of_pH}
We study a multicomponent mixture of macromolecules which can exist in three different charge states. 
Macromolecules with a maximal positive net charge $+m$ are denoted by $\mathrm{M}^+$, those with maximal 
negative net charge $-m$ by $\mathrm{M}^-$ and neutral macromolecules are denoted by $\mathrm{M}$. We 
also consider water molecules 
$\mathrm{H}_2\mathrm{O}$, hydronium ions $\mathrm{H}_3\mathrm{O}^+$  and 
hydroxide ions~$\mathrm{OH}^-$. We describe both protonation and deprotonation 
of the macromolecules as well as the 
self-ionisation of water with the following chemical reactions
\begin{subequations}\label{eq:reactions}
\begin{eqnarray}
 \mathrm{M}^- +m\, \mathrm{H}_3\mathrm{O}^+&\rightleftharpoons & \mathrm{M} + 
m\,\mathrm{H}_2\mathrm{O}\label{eq:neutral_reaction} \quad , 
\\ 
\mathrm{M} + m\, \mathrm{H}_3\mathrm{O}^+ &\rightleftharpoons & \mathrm{M}^+ + 
m\,\mathrm{H}_2\mathrm{O}   
\quad , \label{eq:plus_reaction}\\
\mathrm{H}_3\mathrm{O}^+ + \mathrm{OH}^- &\rightleftharpoons & 2\, \mathrm{H}_2\mathrm{O}  \quad .
\label{eq:self_ionization_reaction}
\end{eqnarray}
\end{subequations}
The average charge state of the macromolecules determined from 
reactions~\eqref{eq:reactions} is controlled by the pH of the mixture. The pH is defined 
as~\cite{pH_IUPAC}
\begin{equation}
\text{pH}=-\log_{10} a_{\mathrm{H}^+}  \label{eq:pH_notional}\quad ,
\end{equation}
with the relative activity of the proton~$a_{\mathrm{H}^+}$ given by
\begin{equation}
a_{\mathrm{H}^+}=\mathrm{exp}\left({\frac{\mu_{\mathrm{H}^+}-\mu_{\mathrm{H}^+}^0}{
k_\mathrm { B } T } }\right) \quad , \label{eq:definition_of_activity}
\end{equation}
where  $k_{\mathrm{B}}$ is the Boltzmann constant, $\mu_{\mathrm{H}^+}$ is the 
chemical potential of protons in the 
system and $\mu_{{\mathrm{H}^+}}^0$ denotes the chemical 
potential of protons in a reference state \cite{IUPAC_GREEN}. The definition of pH 
in Eq.~\eqref{eq:pH_notional} refers to the proton activity. Protons in water are typically 
hydrated~\cite{hydrated_excess_proton_in_water,
hydrated_proton_vibrations_in_clusters}. At chemical equilibrium, the 
proton hydration reaction  
$\mathrm{H}^+ + \mathrm{H}_2\mathrm{O} \rightleftharpoons 
\mathrm{H}_3\mathrm{O}^+$, implies the relation 
$\mu_{\mathrm{H}^+}= \mu_{\mathrm{H}_3\mathrm{O}^+} - 
\mu_{\mathrm{H}_2\mathrm{O}}$, where $\mu_{\mathrm{H}_3\mathrm{O}^+}$ is the 
chemical potential of the hydronium ions and $\mu_{\mathrm{H}_2\mathrm{O}}$ is 
the chemical potential of water.
Therefore, the proton activity (Eq.~\eqref{eq:definition_of_activity}) can be 
written as
\begin{equation}
 a_{\mathrm{H}^+}= \exp{\frac{ \Big(\mu_{\mathrm{H}_3\mathrm{O}^+} - 
\mu_{\mathrm{H}_2\mathrm{O}}\Big)-
\left(\mu_{\mathrm{H}_3\mathrm{O}^+}^{0} - \mu^0_{\mathrm{H}_2\mathrm{O}} 
\right)}{k_\mathrm{B} 
T}}
\label{eq:activity_definition_H3O} \, ,
\end{equation}
where the reference chemical 
potentials,~$\mu_{\mathrm{H}_3\mathrm{O}^+}^0$ and 
$\mu_{\mathrm{H}_2\mathrm{O}}^0$ 
define the pH scale. A standard choice for the reference chemical 
potentials are the chemical potential of the hydronium ions $\mu_{\mathrm{H}_3\mathrm{O}^+}^0$ at 
strong dilution evaluated at a standard concentration 
($n_{\mathrm{\mathrm{H}_3\mathrm{O}^+}}^0=1\,\mathrm{M}$) and the chemical potential of pure water 
$\mu_{\mathrm{H}_2\mathrm{O}}^0$~\cite{IUPAC_GREEN}. In the strong dilution limit, the proton 
activity is~$a_{\mathrm{H^+}}\simeq n_{\mathrm{H}_3\mathrm{O}^+}/n_{\mathrm{H}
_3\mathrm{O}^+}^0$, leading to $\mathrm{pH}\simeq -\log_{10}n_{\mathrm{H}_3\mathrm{O}^+}$, 
where $n_{\mathrm{H}_3\mathrm{O}^+}$ denotes the concentration of the 
$\mathrm{H}_3\mathrm{O}^+$ ions (see the Supplementary Material). In this paper we use
Eqs.~\eqref{eq:pH_notional} and \eqref{eq:activity_definition_H3O} to define the pH. Next, we describe a 
thermodynamic framework to quantify the effect of pH on phase separation behaviour in this 
macromolecular 
system.

\section{Thermodynamics of multicomponent 
mixtures}\label{sect:thermodynamics_of_multicomponent_mixtures}
We consider an incompressible multicomponent mixture in the ($T,P,N_i$) ensemble
with temperature $T$, pressure $P$ and $N_i$ denoting the number of particles of component $i$ in the 
mixture. The corresponding thermodynamic potential is the Gibbs free energy $G(T,P,N_i)$. The chemical 
potential is defined by $\mu_i=\left.\partial G/\partial N_i\right\vert_{T,P,N_{j\ne i}}$, the entropy is
$S=-\left.\partial G/\partial T\right\vert_{P,N_{i}}$ and the  volume of the system is
$V=\left.\partial G/\partial P\right\vert_{T,N_i}$. By 
incompressibility we mean that the molecular volumes of each component, 
$v_i=\left.\partial V/\partial 
N_i\right\vert_{T,P,N_j\ne i}$ are independent of pressure and composition. The volume 
density of the Gibbs free energy is given by~$g(T,P,n_i)=G(T,P,N_i)/V(N_i)$, where we have 
introduced the concentrations~$n_i= N_i/V$ and the volume $V(N_{i})=\sum_i v_i N_i$. 
The chemical potentials can then be calculated from the Gibbs free energy density by
\begin{equation}
\mu_i=v_i\left (g-\sum_k \frac{\partial g}{\partial n_k}n_k\right 
)+\frac{\partial g}{\partial n_i}  \label{eq:chem_pot_g_generic}
\quad .
\end{equation}
We study the multicomponent mixture using a Flory-Huggins mean field free energy model where the 
Gibbs free energy density reads~\cite{huggins41,flory1942thermodynamics}:
\begin{equation}
g=k_\mathrm{B}  T \sum_{k}n_k \ln (n_k v_k) +\sum_{k}w_k 
n_k+\sum_{kl}\frac{\Lambda_{kl}}{2}n_k n_l +P \quad .
\label{eq:g}
\end{equation}
The logarithmic terms stem from the mixing entropy,~$w_k$ denote internal 
free energies of molecules of type $k$ and the interaction parameters $\Lambda_{kl}$ describe the 
contribution to the free energy due to molecular interactions. Molecular interactions can outcompete 
the mixing entropy and cause the emergence of coexisting phases.
Using the free energy density Eq.~\eqref{eq:g}, the chemical potentials are
\begin{equation}
\mu_i=v_i(P-\Sigma)+w_i
+k_\mathrm{B}  T (\ln (n_i v_i)+1)+\sum_k \Lambda_{ik}n_k \quad , \label{eq:chemical_potentials}
\end{equation}
where~$\Sigma$ is defined by 
\begin{equation}
\Sigma=\sum_{kl}\frac{\Lambda_{kl}}{2}n_k n_l+k_\mathrm{B}  T \sum_k n_k \quad . 
\label{eq:sigma_pressure}
\end{equation}

In the multicomponent mixture we consider, the indices $i$, $k$ and $l$ run over the six components 
of the system, which are the three charge states of the macromolecules 
$\mathrm{M,M^+,M^-}$ as well as the three charge states of water $\mathrm{H_2O,H_3O^+ and \ OH^-}$. We 
further consider the molecular volumes of the macromolecules to be all equal, $v=v_\mathrm{M} = 
v_{{\mathrm{M}}^+} = v_{{\mathrm{M}}^-}$, where we have introduced the 
macromolecular volume $v$. We also consider the molecular volumes of water 
and water ions to be the same, $v_0=v_\mathrm{H_2O}=v_\mathrm{H_3O^+}=v_{\mathrm{OH^-}}$, and 
denote them by $v_0$. In order to have a minimal number of components, the 
presence of salt and counterions which neutralise our solution are taken into account implicitly. 
The presence of salt and counterions screen the electrostatic interactions, thus providing a 
characteristic length scale of the interaction potentials between charged species, while the 
counterions mediate the interactions between oppositely charged macromolecules, hence giving an 
effective interaction between them. The effective interactions in our system are captured by the 
interaction matrix $\Lambda_{ij}$. We consider all effective interactions $\Lambda_{ij}=0$ except for 
those between positively and negatively charged macromolecules which we 
choose as $\Lambda_{\mathrm{M}^- \mathrm{M}^+}=\Lambda_{\mathrm{M}^+ 
\mathrm{M}^-}=v\chi_{\mathrm{e}}/\epsilon$ and the effective interaction between neutral macromolecules 
given by $\Lambda_{\mathrm{MM}} =2 
v \chi_{\mathrm{n}}/\epsilon$. Here we have introduced the molecular volumes ratio 
$\epsilon=v_{0}/v$ as well as $\chi_{\mathrm{e}}$ and~$\chi_{\mathrm{n}}$ which 
are interaction parameters characterising the strength of charge-charge and 
neutral-neutral interactions respectively, with this choice, the interaction parameters describe 
the scale corresponding to a water molecule. Attractive interactions are described by negative 
values of these interaction parameters, which we will vary in our study of phase behaviour. In the 
following, we study the system at chemical equilibrium for a fixed pH.

\section{Chemical equilibrium at fixed pH}\label{sect:chem_eq_at_fixed_pH}
We start by stating the conservation laws for a system undergoing 
reactions~\eqref{eq:reactions} and identify the conserved quantities as independent composition 
variables at chemical equilibrium. We then use these independent composition variables to identify 
the thermodynamic conjugated variables at chemical equilibrium. We finalise the section by 
constructing a thermodynamic potential which describes the system at a fixed pH value.

\subsection{Chemical conservation laws}
We consider a system with $s$ different molecular species. If there are $r$ different chemical 
reactions taking place, there exist $c=s-r$ independent composition 
variables. Here $s=6$, the number of independent reactions 
is~$r=3$, therefore the number of independent composition variables is~$c=3$. These independent 
composition variables can be chosen as conserved quantities during chemical reaction events. 
Although the number of independent composition variables is fixed, there is no unique choice of 
conserved variables~\cite{alberty1994legendre}. We choose to use the total number of 
macromolecules in the system $N$, the amount of oxygen $N_{s}$ and the net charge involved in 
the chemical reactions $N_q$ (see the Supplementary Material), given as:
\begin{subequations}\label{eq:conserved_quantities}
\begin{eqnarray}
 N &=& N_{\mathrm{M}^-}+N_\mathrm{M} +N_{\mathrm{M}^+}
\label{eq:total_number_of_macromolecules} \quad , \\ 
  N_{s} &=& N_{\mathrm{H}_{3}\mathrm{O}^+}+N_{\mathrm{OH}^-}+N_{\mathrm{H}_2\mathrm{O}} 
\label{eq:total_solvent_components}
 \quad , \\ N_{q}  &=& 
N_{\mathrm{H}_{3}\mathrm{O}^+}-N_{\mathrm{OH}^-}+m(N_{\mathrm{M}^+}-N_{\mathrm{M}^-}) 
\label{eq:total_number_of_charges} \quad 
.
 \end{eqnarray}
\end{subequations}
Note that the net charge is neutralised by counterions that are not explicitly considered in the 
simplified model.
\subsection{Conjugate thermodynamic variables at chemical equilibrium}

We obtain conditions for chemical equilibrium in terms of the conserved 
variables defined in Eqs.~\eqref{eq:conserved_quantities}, and to do so, we use the variable 
transformation $(N_\mathrm{M},N_\mathrm{H_3O^+},N_\mathrm{H_2O}) \to (N,N_s,N_q)$ to eliminate 
$N_\mathrm{M}$, $N_\mathrm{H_3O^+}$ and $N_\mathrm{H_2O}$. The differential of the Gibbs free 
energy is given by $ dG=-S dT + VdP + \sum_{i} \mu_i dN_i$, which after this variable 
transformation becomes:
\begin{eqnarray}
 dG &=& -S\, dT+V \, dP + \mu_\mathrm{M} \, dN  + 
\mu_{\mathrm{H}_2\mathrm{O}} \, dN_{s} \nonumber \\&& +
(\mu_{H_3O^{+}}-\mu_{\mathrm{H}_2\mathrm{O}}) dN_{q}  \nonumber 
\\&&+(\mu_{\mathrm{M}^-}+m\,\mu_{\mathrm{H}_3\mathrm{O}^+} - 
\mu_\mathrm{M} -m\,\mu_{\mathrm{H}_2\mathrm{O}}) \, 
dN_{\mathrm{M}^-} \nonumber  
\\&&+(\mu_{\mathrm{M}^+} +m\,\mu_{\mathrm{H}_2\mathrm{O}} 
\label{eq:dG_non_chem_eq}
-\mu_\mathrm{M} -m\,\mu_{\mathrm{H}_3\mathrm{O}^+}) \,  dN_{\mathrm{M}^+}  \\&&
+(\mu_{\mathrm{H}_3\mathrm{O}^{+}}+\mu_{\mathrm{OH}^-} -2\mu_{\mathrm{H} 
_2\mathrm { O}}) \, dN_{\mathrm{OH}^-}   
\nonumber 
 \quad .
\end{eqnarray}
If the system has reached chemical equilibrium, the variations $dG$ with respect 
to changes in the non-conserved composition variables $N_\mathrm{M^+}$, $N_\mathrm{M^-}$ and 
$N_\mathrm{OH^-}$ must vanish. This then leads to the following chemical equilibrium 
conditions~\cite{Callen_book}:
\begin{subequations}\label{eq:chem_eq}
\begin{eqnarray}
\mu_{\mathrm{M}^-}+m \,\mu_{\mathrm{H}_{3}\mathrm{O}^+}&=&\mu_{\mathrm{M}} + 
m\, 
 \mu_{\mathrm{H}_2\mathrm{O}} 
\label{eq:relationsmm} \quad , \\
\mu_\mathrm{M}+m \,\mu_{\mathrm{H}_{3}\mathrm{O}^+}&=&\mu_{\mathrm{M}^+}+ m \, 
\mu_{\mathrm{H}_2\mathrm{O}}  
\label{eq:relationsmp}  \quad 
, \\ 
\mu_{\mathrm{H}_3\mathrm{O}^+}+\mu_{\mathrm{OH}^-} &=& 2 \, 
\mu_{\mathrm{H}_2\mathrm{O}}  
\label{eq:self_ionization_of_water} \quad .
\end{eqnarray}
\end{subequations}
Using the chemical equilibrium conditions, the differential of the Gibbs free energy at chemical 
equilibrium is therefore
\begin{eqnarray}
 dG &=& -S\, dT+V \, dP + \mu_\mathrm{M} \, dN  + 
\mu_{\mathrm{H}_2\mathrm{O}} \, dN_{s} \nonumber \\&& +
(\mu_{H_3O^{+}}-\mu_{\mathrm{H}_2\mathrm{O}}) dN_{q}   
\label{eq:dG_chem_eq} \quad ,
 \end{eqnarray}
which explicitly shows that the Gibbs free energy at chemical equilibrium has the dependence 
$G(T,P,N,N_s,N_q)$. This allows us to identify pairs of conjugate thermodynamic variables. From 
Eq.~\eqref{eq:dG_chem_eq} we identify the conjugate thermodynamic variables to the composition 
variables $(N,N_s,N_q)$ as $(\mu_\mathrm{M}, \mu_{\mathrm{H}_2\mathrm{O}}, 
\mu_{H_3O^{+}}-\mu_{\mathrm{H}_2\mathrm{O}})$ respectively. These conjugate variables have to be 
used to obtain Legendre transforms of the thermodynamic potentials~\cite{alberty1994legendre}.

\subsection{Thermodynamic ensemble for fixed pH}
In order to describe the system in an ensemble with pH as a variable, we 
perform a Legendre transform to construct a thermodynamic potential which depends 
on~$\mu_{\mathrm{H}_3\mathrm{O}^+}-\mu_{\mathrm{H}_2\mathrm{O}}$. This 
thermodynamic potential is given by the 
following Legendre transform 
\begin{equation}
 \bar G(T,P,N,N_s,\mu_{\mathrm{H}_3\mathrm{O}^+}-\mu_{\mathrm{H}_2\mathrm{O}}) 
=G-(\mu_{\mathrm{H}_3\mathrm{O}^+}-\mu_{\mathrm{H}_2\mathrm{O}}) N_{q} \quad .
\label{eq:legendre_transform_pH}
\end{equation}
The differential of $\bar G$ reads
\begin{eqnarray}
 d \bar G &=&-SdT+VdP+\mu_{\mathrm{M}} \, dN + 
\mu_{\mathrm{\mathrm{H}_2\mathrm{O}}} \, dN_s
\nonumber \\&& -N_{q}\, 
d(\mu_{\mathrm{H}_3\mathrm{O}^+}-\mu_{\mathrm{H}_2\mathrm{O}}) \quad .
\label{eq:fixed_pH_g}
\end{eqnarray}
We now clarify why fixing this chemical potential difference and the temperature sets the 
pH value of the system. Using Eq.~\eqref{eq:pH_notional} and 
Eq.~\eqref{eq:activity_definition_H3O}, we can express the pH as
\begin{equation}
\mathrm{pH}= \frac{\big( \mu_{\mathrm{H}_3\mathrm{O}^+} - 
\mu_{\mathrm{H}_2\mathrm{O}} \big) - \big( 
\mu_{\mathrm{H}_3\mathrm{O}^+}^{0} - \mu^0_{\mathrm{H}_2\mathrm{O}} \big)
}{k_\mathrm{B} T} \log_{10}e
\label{eq:pH}\quad ,
\end{equation} 
where it is explicitly shown that the pH of the system is set by the relative chemical potential 
of hydronium ions with respect to water, $\mu_{\mathrm{H}_3\mathrm{O}^+} 
- \mu_{\mathrm{H}_2\mathrm{O}}$, and the temperature $T$ of the system. We can also 
define the corresponding free energy in the isochoric ensemble as
\begin{equation}
\bar F=G-(\mu_{\mathrm{H}_3\mathrm{O}^+}-\mu_{\mathrm{H}_2\mathrm{O}}) N_{q}-PV \quad ,
\label{eq:free_energy_F}
\end{equation}
 which has the following differential form
\begin{eqnarray}
 d\bar F&=&-SdT-P dV+ \mu_{\mathrm{M}} \, dN 
 + \mu_{\mathrm{\mathrm{H}_2\mathrm{O}}} \, dN_s
 \nonumber \\&&- N_{q} \, 
d(\mu_{\mathrm{H}_3\mathrm{O}^+}-\mu_{\mathrm{H}_2\mathrm{O}}) \quad .
\label{eq:fixed_pH_fixed_V_diff_compressible_f}
\end{eqnarray}
In our system, the volume~$V$ can be expressed in terms of the 
conserved quantities as $V=vN+v_{0}N_s$, leading to 
$dN_s=dV/v_0-vdN/v_0$. We can then reduce the number of independent variables and 
rewrite the differential form of $\bar F$ as 
\begin{eqnarray}
d\bar F&=&-S\,dT-\Pi \,dV +\bar \mu_{\mathrm{M}} \,dN - N_{q}\,d \bar 
\mu_{\mathrm{H}_3\mathrm{O}^+}  \quad ,
\label{eq:fixed_pH_fixed_V_diff_incompressible}
\end{eqnarray}
where we have introduced the exchange chemical potentials $\bar \mu_{\mathrm{M}}$ of neutral 
macromolecules and $\bar \mu_{\mathrm{H_3O^+}}$ of hydronium ions as well as the osmotic pressure 
$\Pi$, which are defined by
\begin{eqnarray}
 \bar \mu_i &=& \mu_i-\frac{v_i}{v_0}\mu_{\mathrm{H}_2\mathrm{O}} \quad , 
\label{eq:exchange_chem_pot} \\
 \Pi &=& P -\frac{\mu_{\mathrm{H}_2\mathrm{O}}}{v_{0}} \quad ,
\label{eq:osmotic_pressure}
\end{eqnarray}
where $i=\mathrm{M} \ \mathrm{or} \ \mathrm{H_3O^+}$. The free energy $\bar F(T,V,N,\bar 
\mu_{\mathrm{H}_3\mathrm{O}^+})$ depends only on the 
temperature $T$, the volume $V$, the total macromolecule particle number $N$ and the exchange 
chemical potential of the hydrogen ions $\bar \mu_{\mathrm{H}_3\mathrm{O}^+}$. In addition to 
introducing the corresponding thermodynamic potential $\bar F$ which describes an incompressible 
system with a fixed pH value, we have reduced our multicomponent system description from six 
components undergoing three independent chemical reactions to an effective binary mixture with the 
total macromolecule density $n=N/V$ as the only relevant composition variable. We can now ask how the pH 
affects phase separation.

\section{Control of phase separation by pH}\label{sect:control_of_phase_separation_by_pH}
In the following, we discuss how the pH controls both, chemical and phase equilibrium in our 
system. We start by discussing the chemical equilibrium conditions~\eqref{eq:chem_eq} in terms of 
newly defined composition variables and thermodynamic field. After discussing the chemical 
equilibrium conditions in terms of the pH and the newly defined fields, we provide the conditions 
for  phase equilibrium for a system described by the corresponding thermodynamic 
potential defined by~\eqref{eq:free_energy_F}. We end the section by showing a construction of the 
coexisting phases for a given choice of parameters.

\subsection{Thermodynamic fields controlling chemical equilibrium}
We are now interested in discussing how the pH and other parameters influence the
chemical equilibrium described by conditions~\eqref{eq:chem_eq}. In order to do so, we introduce 
thermodynamic fields and composition variables which allow us to interpret the chemical equilibrium 
conditions intuitively. Let us first introduce the following composition variables 
\begin{eqnarray}
\label{eq:n}
n &=& n_{\mathrm{M}^+}+n_{\mathrm{M}^-}+n_{\mathrm{M}} \quad , \\
\label{eq:phi}
\phi &=& \frac{n_{\mathrm{M}^+}+n_{\mathrm{M}^-}}{2n}  \quad ,\\
\label{eq:psi}
\psi &=& \frac{n_{\mathrm{M}^+}-n_{\mathrm{M}^-}}{2n} \quad .
\end{eqnarray}
These composition variables express the total concentration of macromolecules $n$, the fraction of 
charged macromolecules $\phi$ and the difference between concentrations of oppositely charged 
macromolecules relative to the total number of macromolecules $\psi$. 

It is convenient to make a rearrangement of the chemical equilibrium 
conditions~\eqref{eq:chem_eq} as follows
\begin{subequations}\label{eq:chem_eq_pH}
\begin{eqnarray}
\mu_{\mathrm{M}^+}+\mu_{\mathrm{M}^-}   &=& 2 \, \mu_\mathrm{M} 
\label{eq:rel_chem_eq_h_phi} \quad , \\
\mu_{\mathrm{M}^+}-\mu_{\mathrm{M}^-}  &=&  2m \,  
(\mu_{\mathrm{H}_{3}\mathrm{O}^+}-\mu_{\mathrm{H}_{2}\mathrm{O}})
\label{eq:rel_chem_eq_h_psi} \quad , \\
\mu_{\mathrm{H}_{2}\mathrm{O}} - \mu_{\mathrm{OH}^-}   &=&  
\mu_{\mathrm{H}_{3}\mathrm{O}^+}-\mu_{\mathrm{H}_{2}\mathrm{O}}
\label{eq:rel_chem_pot_self_ion_water} \quad ,
\end{eqnarray}
\end{subequations}
where we see that the right hand sides of Eqs.~\eqref{eq:rel_chem_eq_h_psi} and 
\eqref{eq:rel_chem_pot_self_ion_water} are determined by the pH. We now write the 
conditions~\eqref{eq:rel_chem_eq_h_phi} and \eqref{eq:rel_chem_eq_h_psi} in terms 
of the composition variables using the expressions of the chemical potentials 
(Eqs.~\eqref{eq:chemical_potentials_of_all}) leading to
\begin{eqnarray}
k_\mathrm{B}T\ln \frac{\phi^2-\psi^2}{(1-2\phi)^2}+\frac{2 {v} \chi_{\mathrm{e}} n 
\phi}{\epsilon} -\frac{4{v} \chi_{\mathrm{n}} 
n(1-2\phi)}{\epsilon}  = h_\phi \quad  ,
\label{eq:phi2psi2} \\
 k_\mathrm{B}T\ln \frac{\phi+\psi}{\phi-\psi}-\frac{2 {v}\chi_{\mathrm{e}} n 
\psi}{\epsilon}  
= h_\psi  
\label{eq:phipsi} 
\quad ,
\end{eqnarray}
where we have defined 
\begin{eqnarray}
 h_\phi&=& 2w_\mathrm{M}-w_{\mathrm{M}^+}-w_{\mathrm{M}^-}\quad , 
\label{eq:h_phi}\\
 h_\psi&=&2m\,( 
\mu_{\mathrm{H}_{3}\mathrm{O}^+}-\mu_{\mathrm{H}_{2}\mathrm{O}})-w_{\mathrm{M}^+}+w_{\mathrm{M}^-
} \quad . 
\label{eq:h_psi}
\end{eqnarray}
 These quantities play the role of fields controlling the chemical equilibrium and phase separation 
behaviour of the system. The molecular field~$h_\phi$ characterises which of the 
macromolecular charge states is energetically favoured due to their internal free energies~$w_i$, 
while the field~$h_\psi$ expresses deviations of the pH from its value at the isoelectric point (pI) of 
the macromolecules. Remember that the isoelectric point is the value of the pH for 
which macromolecules are on average neutral. Thus, in our system, the isoelectric point is 
defined as the value $\mathrm{pI}$ of the $\mathrm{pH}$ at which the charged macromolecules obey, 
$n_{\mathrm{M^+}}=n_{\mathrm{M^-}}$, or equivalently $\psi=0$, which implies $h_\psi=0$ 
via~Eq.~\eqref{eq:phipsi}. Using Eq.~\eqref{eq:pH}, we find an expression for the pI value:
\begin{equation}
 \text{pI}=\left(\frac{w_{\mathrm{M}^+}-w_{\mathrm{M}^-}}{2mk_\mathrm{B} 
T}-\frac{\mu_{\mathrm{H}_{3}\mathrm{O}^+}^0-\mu_{\mathrm{H}_{2}\mathrm{O}}^0}{
k_\mathrm{B} T} \right)\log_{10} e 
\label{eq:pI} \quad .
\end{equation}
We can therefore express the field $h_\psi$ in terms of the pH and the pI as follows
\begin{equation}
 \frac{h_\psi} {k_\mathrm{B}T}= \frac{2m}{\log_{10}e}(\text{pI}-\text{pH}) \label{eq:hpsi_pH} \quad 
,
\end{equation}
which explicitly shows how $h_\psi$ characterises deviations of the system from its 
isoelectric point. For given~$h_\phi$ and $h_\psi$, the composition variables~$\phi$ and~$\psi$
can be determined from Eqs.~\eqref{eq:phi2psi2}-\eqref{eq:phipsi} as a function of the total 
macromolecule density $n$, the temperature $T$ and the pH. One symmetry can be identified in 
Eqs.~\eqref{eq:rel_chem_eq_h_phi} and \eqref{eq:rel_chem_eq_h_psi}, namely that the system behaves 
identically under the transformation $\psi \to -\psi$ and $h_\psi \to -h_\psi$, which will be 
reflected in the phase diagrams as a function of deviations from the isoelectric point. This 
symmetry stems from both, considering that positively and negatively charged macromolecules have 
the same interaction with the rest of the components as well as from choosing their molecular 
volumes to be the same. Changing any of the two previously mentioned conditions would break this 
symmetry. We must note that in a more realistic scenario there would be differences in solvation of 
the two different charged states of the macromolecule \cite{negative_charged_molecules_solvation}. 

We find the concentrations of water components by using the chemical equilibrium condition corresponding 
to the self-ionisation of water reaction and Eqs.~\eqref{eq:mu_rel_hydroxide}-\eqref{eq:mu_rel_water}. 
We then express the concentrations $n_{\mathrm{H_3O^+}}$, $n_{\mathrm{OH^-}}$ and $n_{\mathrm{H_2O}}$ in 
terms of two new fields $h_H$ and $h_O$ which only depend on the molecular properties of the water 
components and the pH of the system (the definitions of these 
fields as well as explicit formulas for the concentration of water components are provided in the 
Supplementary Material).

We have shown that chemical equilibrium can be fully accounted for by the internal 
free energies of all species $w_i$ which we take to be constant, the temperature $T$, the pH (or 
equivalently the chemical potential difference $\mu_{\mathrm{H}_{3}\mathrm{O}^+} - 
\mu_{\mathrm{H}_{2}\mathrm{O}}$) and the total macromolecule concentration $n$.

\subsection{Phase coexistence in the pH ensemble} 
We are interested in describing the phase behaviour of the system in the pH ensemble. To this end we 
 make use of the composition variables $n$, $\phi$ and $\psi$ and the fields $h_\phi$, 
$h_\psi$, $h_H$ and $h_O$ defined in the previous section. 
Using Eq.~\eqref{eq:free_energy_F} we define the free energy density $\bar f(T,n,\bar 
\mu_{\mathrm{H}_3\mathrm{O}^+})=\bar F(T,V,N,\bar \mu_{\mathrm{H}_3\mathrm{O}^+})/V$ which reads
\begin{eqnarray}
\bar f(T,n,\bar \mu_{\mathrm{H_3O^+}}) &=&k_\mathrm{B}  T  \bigg[
n(\phi+\psi)\ln(v n(\phi+\psi))+ n(\phi-\psi)\ln(v 
n(\phi-\psi)) 
\nonumber \\&&+ n(1-2\phi)\ln(v n(1-2\phi)) 
\nonumber
\\&&+\frac{1}{v_0}(1-vn) 
\ln\left(\frac{1-vn}{1+e^{h_{\mathrm{H}}/k_\mathrm{B}T+h_{\psi}/2mk_\mathrm{B}T}+e^{h_{\mathrm
{O}}/k_\mathrm{B}T -h_\psi/2mk_\mathrm{B}T } } \right)\bigg]\nonumber \\&&+
\frac{v}{\epsilon}  
\chi_\mathrm{e}  n^2 
(\phi^2-\psi^2)  
   +  \frac{v}{\epsilon} \chi_\mathrm{n} n^2(1-2\phi)^2 - h_\phi n\phi 
-h_{\psi} n\psi   \nonumber \\&&+ w_\mathrm{M} n 
+\frac{w_{\mathrm{H}_2\mathrm{O}}}{v_0}\big(1-v n \big  
) \quad \label{eq:free_energy_density_f}  ,
\end{eqnarray}
where the functions $\phi(T,n,\bar 
\mu_{\mathrm{H}_3\mathrm{O}^+})$ and $\psi(T,n,\bar 
\mu_{\mathrm{H}_3\mathrm{O}^+})$ are defined implicitly in Eqs.~\eqref{eq:phi2psi2} and 
\eqref{eq:phipsi} in terms of the temperature $T$, the exchange chemical potential of 
hydronium ions $\bar \mu_\mathrm{H_3O^+}$ and the total macromolecule density $n$.

\begin{figure}
\includegraphics[width=\textwidth]{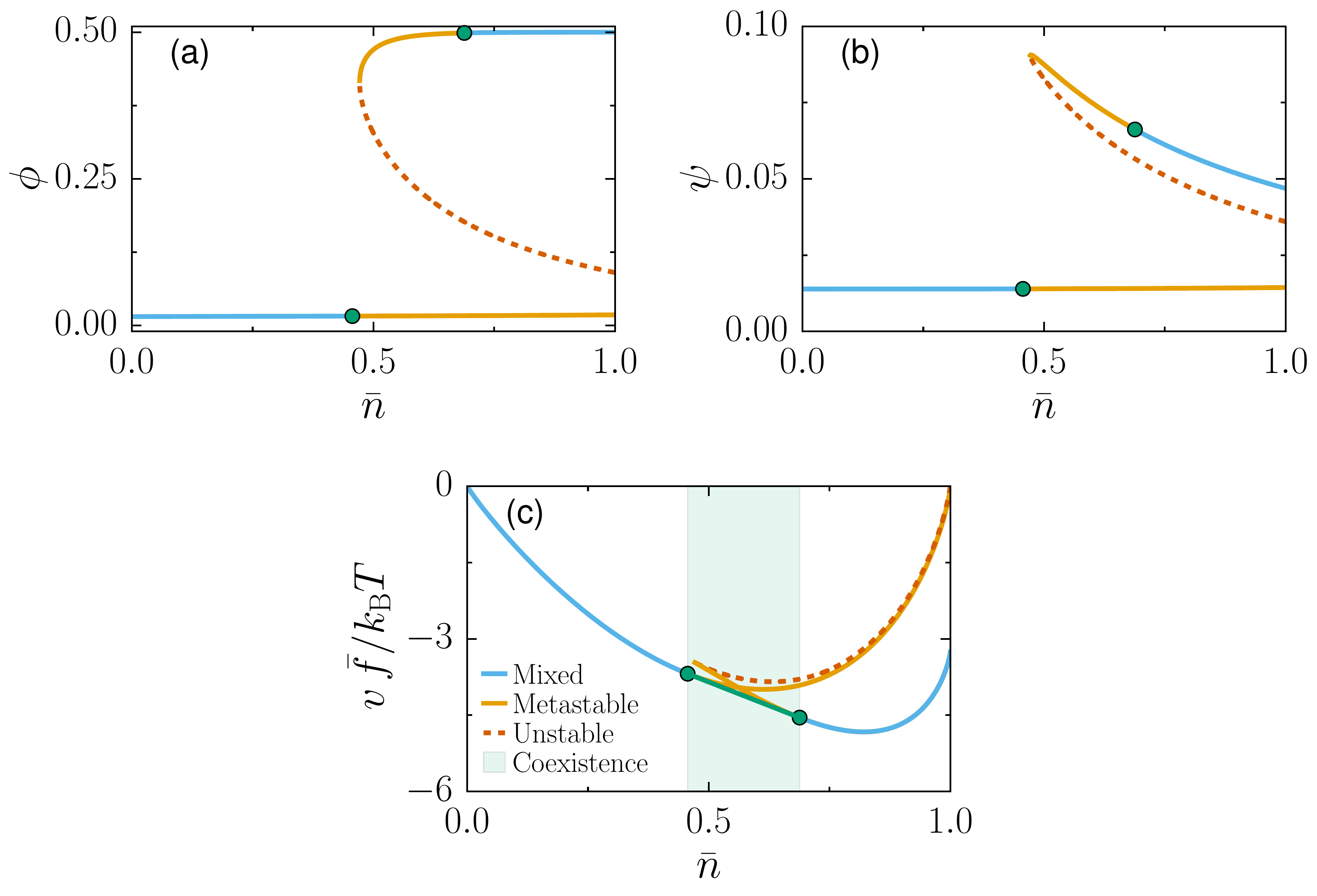}
\caption{Chemical equilibrium conditions and free energy density at chemical equilibrium. Multiple 
solutions are found for $\phi$ (a) and $\psi$ (b) as a function of the total macromolecule volume 
fraction $\bar n$, enabling the system to exhibit phase separation between different branches 
of the chemical equilibrium. The blue solid lines correspond to equilibrium concentrations where 
the system remains homogeneous, the orange solid lines represent solutions to the chemical 
equilibrium relations which are metastable states and the dotted red line shows the unstable states. (c) 
Maxwell construction for the dimensionless free energy density $v \bar f/k_\mathrm{B}T$ as a function of 
the total macromolecule volume fraction, the green line describes the region of macromolecule volume 
fraction where the system split into two phases with different compositions given by the green circles. 
Parameters $\chi_\mathrm{e}/k_\mathrm{B}T= -3$, $\chi_\mathrm{n}= 0$, $\mathrm{pI}-\mathrm{pH}=0.2$, 
$h_\phi/k_\mathrm{B}T = -10$ and $\epsilon=0.1$, apply to all panels.}
\label{fig:law_of_mass_action_common_tangent}
\end{figure}

We now discuss the phase coexistence conditions for this incompressible system at fixed temperature 
$T$ and at a fixed pH value, which can be obtained by a Maxwell construction. It follows from 
Eq.~\eqref{eq:fixed_pH_fixed_V_diff_incompressible} 
that the exchange chemical potential of neutral macromolecules is given by~$\bar 
\mu_\mathrm{M}=\left. \partial \bar f/ \partial n\right|_{T,\bar \mu_{\mathrm{H}_3\mathrm{O}^+}}$ 
and that the osmotic pressure is given by $\Pi=\bar f-n\left. \partial \bar f/ \partial 
n\right|_{T,\bar \mu_{\mathrm{H}_3\mathrm{O}^+}}$. Using the free energy density~$\bar f$, we write 
the phase equilibrium conditions describing equal exchange chemical potentials of the neutral 
macromolecules and equal osmotic pressures in both phases~\cite{safran_book}:
\begin{subequations}\label{eq:coexistence_conditions}
\begin{eqnarray}
 \bar \mu_{\mathrm{M}}(n^{\text{I}})&=&\bar 
\mu_{\mathrm{M}}(n^{\text{II}}) \quad , \label{eq:same_chem_pot} \\
 \bar \mu_{\mathrm{M}}(n^{\text{I}})
&=& \frac{\bar f(n^{\text{II}})-\bar f(n^{\text{I}})}{n^{\text{II}}-n^{\text{I}}} \quad ,
\label{eq:same_line_coex}
\end{eqnarray}
\end{subequations}
where the superscripts~$\text{I},\text{II}$ denote the two coexisting phases. We did not write 
the explicit dependence of the relative chemical potential $\bar \mu_\mathrm{M}$ and of the free 
energy density $\bar f$ on the temperature $T$ and on the relative chemical potential $\bar 
\mu_\mathrm{H_3O^+}$. These conditions correspond to the common tangent 
construction~\cite{safran_book}. We can find the coexisting phases by 
first calculating the values of $\phi$ and $\psi$ using Eqs.~\eqref{eq:phi2psi2} and 
\eqref{eq:phipsi} (Fig.~\ref{fig:law_of_mass_action_common_tangent}(a,b)), and then by doing a 
common tangent construction for the free energy density Eq.~\eqref{eq:free_energy_density_f} evaluated 
as a function of the total macromolecule volume fraction $\bar n = 
vn$~(Fig.~\ref{fig:law_of_mass_action_common_tangent}(c)). The phase diagrams can then be readily
constructed by repeating the steps described above for different parameter values.

\section{Phase diagrams at the isoelectric point}\label{sect:phase_diagrams_at_pI}
\begin{figure}
 \includegraphics[width=0.83\textwidth]{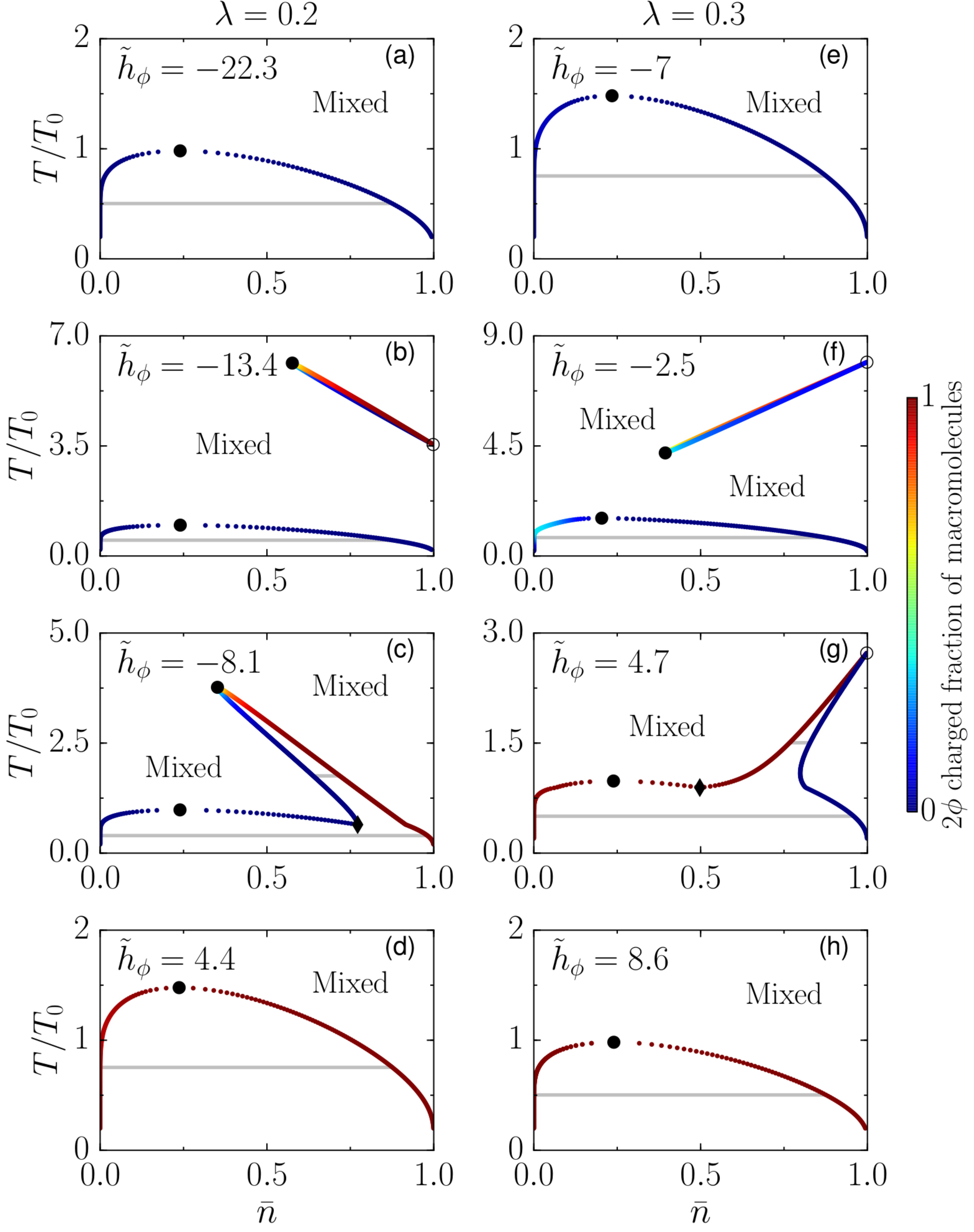}
\centering
 \caption{Topologies of the phase diagrams for varying values of the normalized molecular field $\tilde 
h_\phi=h_\phi/k_\mathrm{B}T_0$ defined in Eq.~\eqref{eq:h_phi} where $T_0$ is a reference temperature. 
(a-d) phase diagrams for a system in which charge-charge interactions are slightly stronger than 
neutral-neutral interactions. (e-h) Phase diagrams for a system in which neutral-neutral interactions 
are slightly stronger than charge-charge interactions. The binodals are given by the coloured points 
which denote coexisting phases. Tie lines (grey solid lines) connect coexisting phases and are 
horizontal. The regions within the binodals undergo a demixing transition, whereas the regions outside 
the binodal lines remain well mixed.  The critical points where phases become indistinguishable are 
denoted by black circles, first order transition points where there is a discontinuity in the value of 
$\phi$ are denoted by white circles and triple points are denoted by black diamonds. A thorough 
explanation of the phase diagrams is given in the main text. Parameters $\chi=-8.5$ and $\epsilon=0.1$, 
apply to all panels. The colorbar indicates the value of the charged fraction of macromolecules $\phi$.}
\label{fig:topologies_phase_diagrams}
\end{figure}

\begin{figure}
 \includegraphics[width=\textwidth]{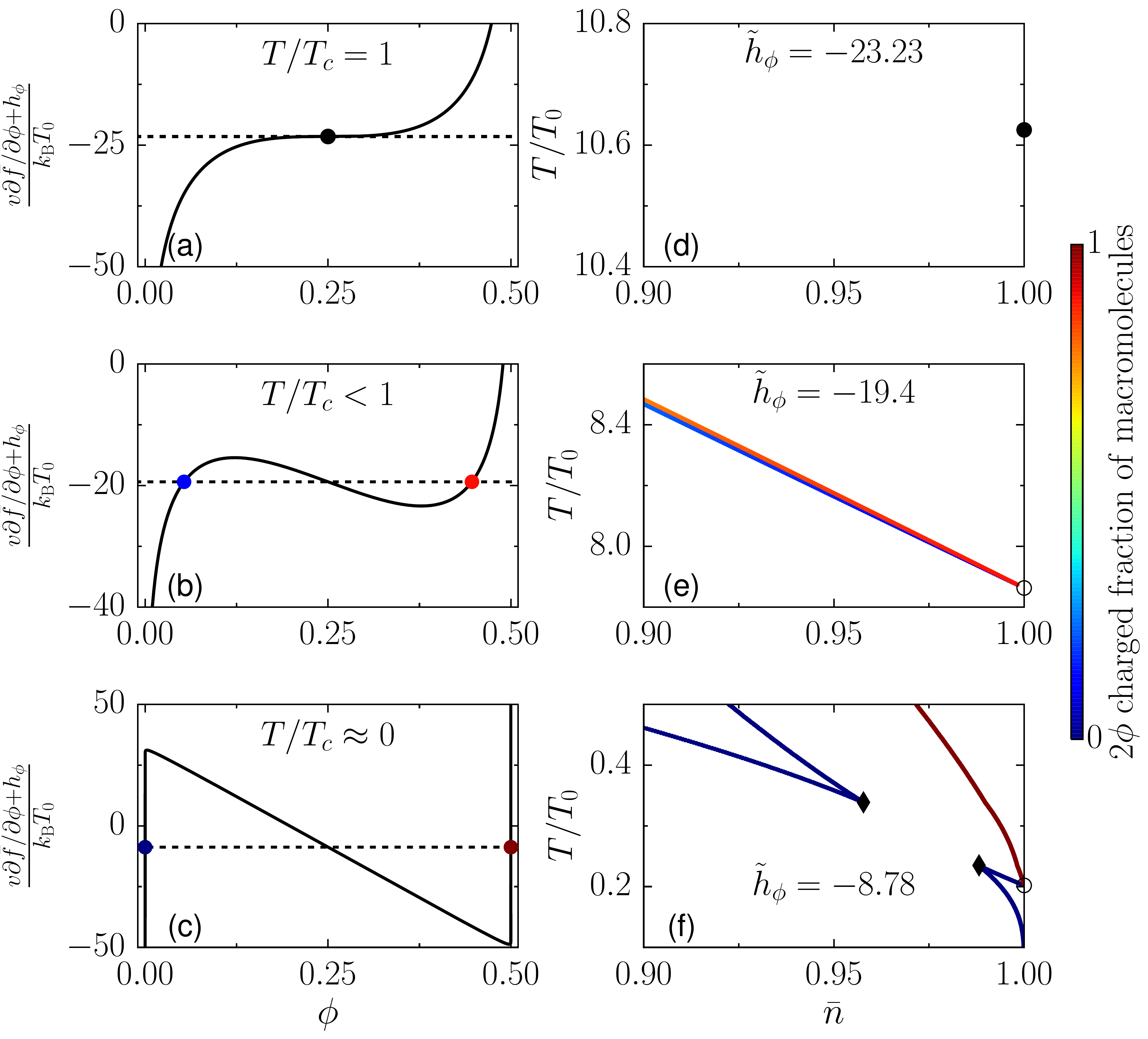}
  \centering
 \caption{Critical behaviour on the $\bar n=1$ line. (a-c) Derivative of the 
free energy with respect to $\phi$ for different temperature values . (a) Inflection point corresponding 
to the critical point defined in Eq.~\eqref{eq:crit_values_at_n_1} shown as a filled black circle. The 
black dotted line shows the value of $\tilde h_\phi=\tilde h_\phi^{c}\simeq -23.23$. (b) Emergence of a 
maximum and a minimum for $T<T_c$, coexisting phases are shown as two colored circles (the color encodes 
their value of $\phi$). (c) Derivative of the free energy density for $T/T_0=0.2$, implying $T/T_c\ll1$. 
 (d-f) Phase diagrams for fixed $\tilde h_\phi = h_\phi/k_\mathrm{B}T_0$ in the vicinity of the 
transition point at $\bar n =1$.(d) The filled black circle is the isolated critical point defined in 
Eq.~\eqref{eq:crit_values_at_n_1} corresponding to (a). (e) The phase coexistence lines shown in blue 
and red end on the $\bar n=1$ line at the first order transition point (open circle) defined in (b). (f) 
The coexistence region connected to the $\bar n=1$ axis, merges with the quasi-binary region, leading to 
the appearance of two triple points (black diamonds in (f)). Parameters $\chi=-8.5$, $\lambda=0.2$ and 
$\epsilon=0.1$, apply to all panels. $T_0$ is a reference temperature with $T_c/T_0 \simeq 10.6$.}
\label{fig:behavior_near_1}
\end{figure}

In this section, we investigate different phase diagrams which can be obtained by varying 
temperature at constant total number of macromolecules, keeping the system at its isoelectric point. 
In order to discuss the effects of temperature  we use a weighted sum of the interaction parameters 
$\chi= \chi_{\mathrm{e}}+4\chi_\mathrm{n}$, the ratio of  interaction parameters 
$\lambda=\frac{2\chi_{\mathrm{n}}}{\chi_{\mathrm{e}}+4\chi_{\mathrm{n}}}$, the molecular field 
$h_\phi$ and rewrite Eq.~\eqref{eq:phi2psi2} and Eq.~\eqref{eq:free_energy_density_f} in the following 
compact way:
\begin{eqnarray}
h_\phi &=& 2k_\mathrm{B}T\ln \frac{\phi}{(1-2\phi)}+ \frac{2 \chi \bar{n}  (\phi 
-\lambda)}{\epsilon}  \quad  , \label{eq:phi2psi0} \\
v \bar f(T,n,\bar \mu_{\mathrm{H_3O^+}}) &=&k_\mathrm{B}  T  \bigg[
2\bar n \phi \ln(\bar n \phi)
+ \bar n(1-2\phi)\ln (\bar n(1-2\phi)) 
\nonumber
\\&&+\frac{1}{\epsilon}(1-\bar n)  
\ln\left(\frac{1-\bar n}{1+e^{h_{\mathrm{H}}/\mathrm{k_\mathrm{B}}T}+e^{h_{\mathrm {O}}/k_\mathrm{B}T}}   
\right) \bigg]\nonumber \\&&+ 
\chi \epsilon^{-1} \bar n^2 
(\phi^2-2\lambda \phi+\lambda/2)   - h_\phi \bar n\phi   \nonumber \\&&+\left( w_\mathrm{M} 
-\frac{w_{\mathrm{H}_2\mathrm{O}}}{\epsilon} \right) \bar n + \frac{w_\mathrm{H_2O}}{\epsilon } \quad 
\label{eq:free_energy_density_vf} 
\quad .
\end{eqnarray}
We further consider attractive interactions $\chi<0$ and $\lambda>0$. In order to construct phase 
diagrams as a function of temperature we rescale all the variables which have energy units with 
$k_\mathrm{B}T_0$, where $T_0$ is a reference temperature. Free energy minimisation at constant $ 
h_\phi ,  \chi , \lambda, \epsilon, h_\mathrm{H}, h_\mathrm{O}$ and $T$ together with a common tangent 
construction, lead to different possible topologies of phase diagrams which are summarised 
in Fig.~\ref{fig:topologies_phase_diagrams} and Fig.~\ref{fig:behavior_near_1}.  

For $2\chi (1/2-\lambda)> \epsilon\,h_{\phi}$, or $ \epsilon \, h_{\phi}>-2\chi\lambda$, the diagram has 
the same topology as that of a simple two component mixture~\cite{rubinstein2003polymer}, see 
Fig.~\ref{fig:topologies_phase_diagrams}(a,d,e,h). At low temperature the system demixes in  a low 
density and a high density phase. For $\epsilon h_{\phi}\ll2\chi (1/2-\lambda)$, the 
proportion of charged molecules is exponentially small and the system behaves like a neutral polymer 
solution. In this case the system separates into a low density phase and a high density phase composed 
essentially of neutral macromolecules $\phi \approx 0$. The coexistence curve is bell shaped and by 
construction the tie lines are parallel to the $\bar n$ axis. The point at which the tangent is parallel 
to this axis is a critical point (Fig.~\ref{fig:topologies_phase_diagrams}(a,e)). It belongs to the same 
universality class as a liquid-vapour critical point. For  $\epsilon h_{\phi}\gg-2\chi\lambda $, the 
concentration of neutral molecules is exponentially small and the system demixes between a low density 
phase and a high density phase composed essentially of charged molecules $\phi \approx 1/2$. Again the 
coexistence curve is "bell shaped" and one observes the existence of an isolated critical point 
(Fig.~\ref{fig:topologies_phase_diagrams}(d,h)). 

In both limits, the mean-field calculation of the critical coordinates can be done analytically. The 
details are given in the Supplementary Material. In these limits, the critical values are
given by: 
\begin{subequations}\label{eq:critical_values_effective_binary}
\begin{eqnarray}
\bar n_c^{b}&=&\frac{\sqrt{\epsilon}}{1+\sqrt{\epsilon}} \quad , \ \text{for} \ \phi 
=0 \ \text{and} \ \phi =\frac{1}{2}  \ , \\  
k T_c^b &=& - \frac{2\chi_{n}}{\left(1+\sqrt{\epsilon} \right)^{2}}  \quad , \ 
\text{for} \ \phi = 0 \ , \label{eq:chi_m_b}
\\
k T_c^b &=& - \frac{\chi_{e}}{2\left(1+\sqrt{\epsilon} \right)^{2}}  \quad , \ \text{for} \ 
\phi 
=\frac{1}{2} \ ,
\label{eq:chi_e_b}  
\end{eqnarray}
\end{subequations}
where we have used the the condition for critical points, $\partial^2 \bar f/\partial 
n^2=0$ and $\partial^3\bar f/\partial n^3=0$. We refer hereafter to these coexistence regions as 
quasi-binary regions. 

For intermediate values of the molecular field, $2\chi (1/2-\lambda)\lesssim 
\epsilon \, h_{\phi}\lesssim-2\chi\lambda $ the possible topologies of phase diagrams are more complex. 
Increasing the value of $h_\phi$, from very large negative values towards positive values, leads to the 
emergence of a second coexistence region (Fig.~\ref{fig:topologies_phase_diagrams}(b,f)). This 
coexistence region, disconnected from the quasi-binary region, is bounded by a critical point and it is
connected to the $\bar n=1$ axis at a single point, where the tie line span vanishes. The transition 
point corresponds to a first order transition on the $\bar n$ line, where $\phi$ undergoes a 
discontinuity. A second coexistence region emerges via a critical point, whose values are given by 
(Fig.~\ref{fig:behavior_near_1}(a,d)):
\begin{subequations}\label{eq:crit_values_at_n_1}
\begin{eqnarray}
 \phi_c &=& \frac{1}{4} \label{eq:phi_crit_n_1} \quad , \label{eq:crit_phi_n_1} \\
 k T_c&=&-\frac{\chi}{8 \epsilon} \quad ,\label{eq:crit_chis_n_1} 
\\
 h_{\phi,c}&=& \frac{\chi  (2+\ln(2)-8 \lambda)}{4\epsilon}
\label{eq:crit_mol_field_n_1} \quad .
\end{eqnarray}
\end{subequations}
The two phase region collapses to one point on the $\bar n=1$ line, both in the first order and in the 
second order scenarios, this results from the existence of only one singularity in $\phi$ on the $\bar 
n=1$ line. Note that $h_{\phi,c}$ can be either positive or negative, so that one can have a critical 
point on the $\bar n=1$ line both in the neutral and in the charged regimes, see the Supplementary 
Material.

For some values of $h_\phi$, with $h_\phi>h_{\phi,c}$, the two coexistence regions 
merge (Fig.~\ref{fig:topologies_phase_diagrams}(c,g)). Depending on $\lambda$, there may be two 
different generic scenarios. We first explain what happens for $\lambda<1/4$. In this case, the two 
regions merge, giving rise to two triple points (Fig.~\ref{fig:behavior_near_1}(f)). The two triple 
points have a low density phase enriched in neutral macromolecules coexisting with both, an intermediate 
phase with a large macromolecule concentration, which is also rich in neutral macromolecules and with a 
macromolecule dense phase of essentially charged macromolecules. For $\epsilon \, 
h_\phi>2\chi(1/4-\lambda)$, one triple point and the first order transition point vanish. This bound for 
$h_\phi$ is found by solving the coexistence conditions at $\bar n=1$, for $T=0$. For increasing values 
of $h_\phi$, the remaining triple point moves towards and eventually merges with the critical point of 
the quasi-binary region, leading to a coexistence region which has only one critical point 
(Fig.~\ref{fig:topologies_phase_diagrams}(c)). Finally, for larger values of $h_\phi$, the system 
behaves as a binary mixture of charged macromolecules and solvent 
(Fig.~\ref{fig:topologies_phase_diagrams}(d)). In contrast, for $\lambda>1/4$ and $h_\phi>h_{\phi,c}$, 
when the two regions merge, there is only one triple point, where two phases which are essentially 
enriched in charged macromolecules, coexist with a high concentration phase enriched in neutral 
macromolecules (Fig.~\ref{fig:topologies_phase_diagrams}(g)). For larger values of $h_\phi$, the triple 
point vanishes together with the first order transition point at $T=0$ and 
$\epsilon\,h_\phi=2\chi(1/4-\lambda)$. This vanishing leads again to the quasi-binary mixture of charged 
macromolecules and solvent (Fig.~\ref{fig:topologies_phase_diagrams}(h)). 

The existence of a transition point on the $\bar n=1$ line leads to the different topologies of the 
phase diagrams~(Fig.~\ref{fig:behavior_near_1}). To understand the behaviour of this point, we analyse 
the derivative of the free energy density with respect to $\phi$ at $\bar n=1$, in particular for 
$\lambda=0.2$ (Fig.~\ref{fig:behavior_near_1}(a-c)).  At the critical temperature, 
Eq.~\eqref{eq:crit_chis_n_1}, there is an inflection point (Fig.~\ref{fig:behavior_near_1}(a)), which 
translates into a critical point for $h_\phi=h_{\phi,c}$ at $T=T_c$ (Fig.~\ref{fig:behavior_near_1}(d)). 
For lower values of the temperature, $T<T_c$ one finds a first order transition point at 
$h_\phi>h_{\phi,{c}}$, in which two phases coexist at $\bar n=1$ (Fig.~\ref{fig:behavior_near_1}(b,e)). 
For lower temperatures, the region which extends from the first order transition point merges with the 
rest of the phase diagram (Fig.~\ref{fig:behavior_near_1}(f)) and the derivative of the free energy 
density becomes increasingly dominated by a linear term in $\phi$, given by 
$2\chi(\phi-\lambda)/\epsilon$ (Fig.~\ref{fig:behavior_near_1}(c)). For $T=0$, there is a corresponding 
value of the molecular field, $\epsilon\,h_\phi=2\chi(1/4-\lambda)$, for which we find a solution to the 
coexistence conditions. For values $\epsilon h_\phi>2\chi(1/4-\lambda)$ there is no longer a transition 
point at $\bar n=1$. We only focus on $\lambda=0.2$ because the behaviour of the transition point 
at $\bar n=1$ is similar for $\lambda>0.25$. 
Now that we developed a detailed understanding of phase diagrams at the isoelectric point, we can use 
the developed framework to study the effects of varying pH.

\section{Phase separation for varying pH}\label{sect:phase_coexistence_away_from_pH}
Here we study how deviations in pH with respect to the isoelectric point affect the phase behaviour 
of the system while keeping the temperature constant. To this end, we study phase diagrams as a function 
of $\mathrm{pI}-\mathrm{pH}$ (Eq.~\eqref{eq:hpsi_pH}) and the total macromolecular volume fraction $\bar 
n$. We construct phase diagrams for different values of the molecular field $h_\phi$, the interaction 
strength among charged macromolecules $\chi_{\mathrm{e}}$ and the interaction strength among neutral 
macromolecules $\chi_{\mathrm{n}}$. Varying the pH in our model leads to very characteristic features in 
the phase diagrams which allow us to distinguish the dominant interaction driving phase separation.

\begin{figure}[t!]
 \centering
\includegraphics[width=\textwidth]{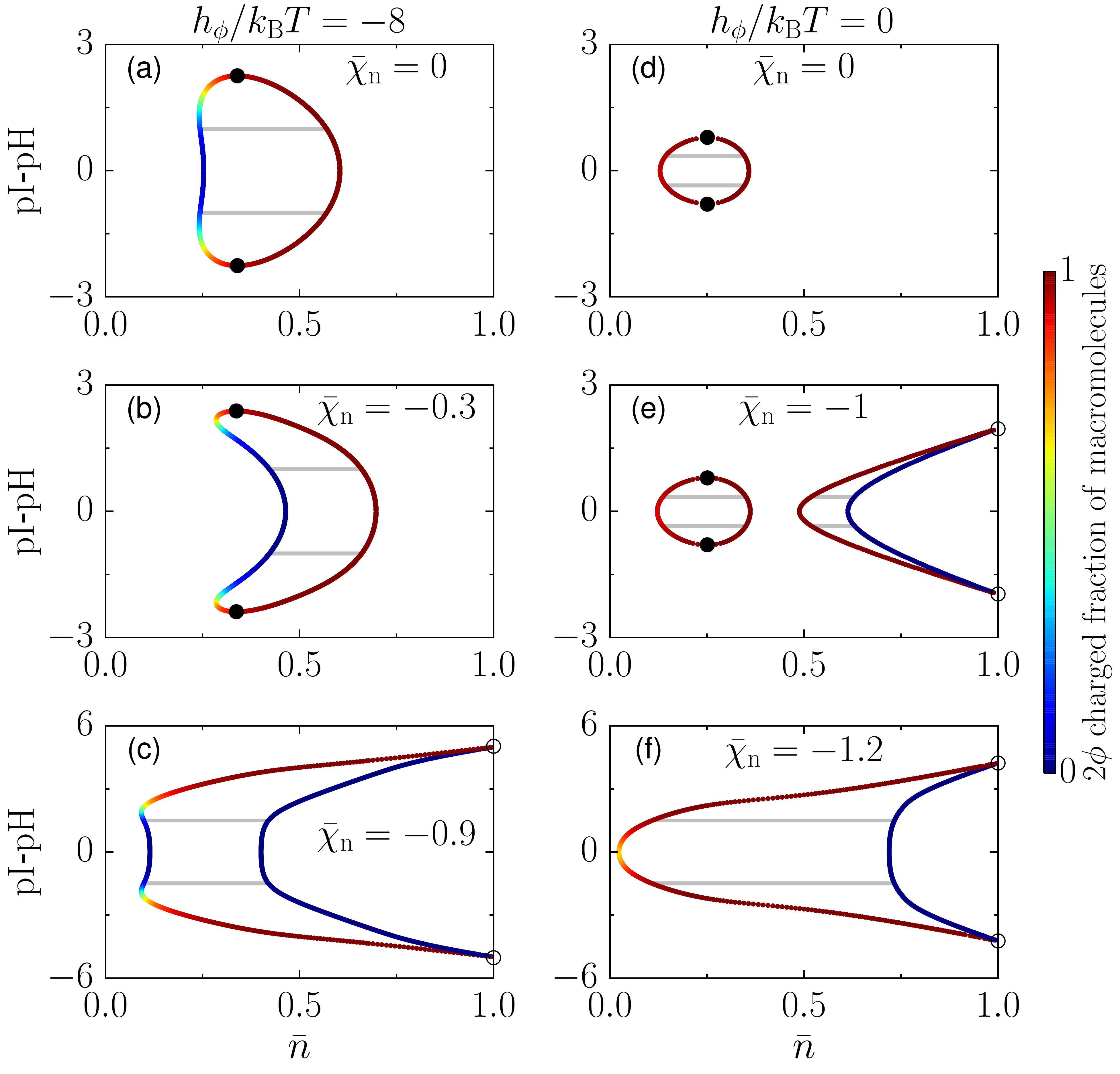}
  \caption{Phase behaviour as a function of pH for fixed interaction 
strength between charges $\chi_\mathrm{e}/k_\mathrm{B}T=-3.5$ and   
varying values of the interaction strength $\bar \chi_{\mathrm{n}} = 
\chi_\mathrm{n}/k_\mathrm{B}T$ between neutral macromolecules. (a-c) Phase diagrams with 
neutral macromolecules energetically favoured ($h_\phi/k_\mathrm{B}T=-8$). (a) in the absence of 
neutral-neutral interactions there is a small region in the diagram where there is reentrant phase 
separation behaviour. (b) Small values of neutral-neutral interactions lead to a reduction of the 
demixing region. (c) Increasing the neutral-neutral attraction even further, the two critical points 
merge and two first order transition points appear at $\bar n=1$, which has a discontinuity $\phi$ and 
$\psi$. (d-f) Phase diagrams with charged macromolecules energetically favoured 
($h_\phi/k_\mathrm{B}T=0$): (d) An effective binary mixture at the pI shows a simple mixing behaviour 
while deviating from the isoelectric point. (e) For large enough interactions between neutral 
macromolecules, a second disconnected region appears, such region ends in two first order transition 
points at $\bar n=1$. (f) Increasing the neutral-neutral 
interactions further, the two regions merge giving rise to a broadening of the demixing region while the 
critical points vanish and the coexistence region connects to the $\bar n=1$ line with two first 
order transition points. Parameters $\epsilon=0.1$ and $m=1$ apply to all panels.}
\label{fig:phase_diagrams_away_from_pI_both_interactions}
\end{figure}
Let us first consider the case where neutral molecules are energetically favoured over charged molecules 
($h_\phi=-8$, Fig.~\ref{fig:phase_diagrams_away_from_pI_both_interactions}(a-c)).
In this case, a system with only charge-charge interactions 
(Fig.~\ref{fig:phase_diagrams_away_from_pI_both_interactions}(a), $\chi_{\mathrm{n}}=0$) 
exhibits reentrant behaviour when changing the pH where the corresponding domain in the phase 
diagram is enclosed by two critical points. Beyond these points, phase separation is not possible for 
any value of the macromolecular volume fraction $\bar n$ while between the critical points, there is a 
range in $\bar n$ where phase separation can occur. The degree of such phase separation is maximal 
at the isoelectric point pH=pI, which is characterised by the largest difference between coexisting 
phases in their macromolecular volume fraction, as well as in their charged fraction between coexisting 
phases. Deviating from the isoelectric point corresponds to lowering the amount of one of the charged 
components ($\psi \not=0$, Eq.~\eqref{eq:psi}). This change in the relative composition between charged 
macromolecules decreases the interaction term among charged components (proportional to $n_+  n_-$) and 
thereby lowers their propensity to phase separate. There is a small range in macromolecular volume 
fractions where phase separation is absent at the isoelectric point but can be triggered by changing the 
pH value away from pI. This range strongly increases for stronger interactions among neutral 
macromolecules  ($\chi_{\mathrm{n}}$ more negative, 
Fig.~\ref{fig:phase_diagrams_away_from_pI_both_interactions}(b)). Such behaviour for phase separation is 
unexpected because phase separation occurs despite of an asymmetric ratio of the charged 
macromolecules. It emerges as a consequence of a reduction in the mixing entropy of the macromolecules 
by moving away from the isoelectric point which in turn allows the system to phase separate at lower 
values of charged fraction of macromolecules $\phi$. Even though the system shows phase separation at 
lower macromolecular volume fraction, the region of phase separation decreases for increasing deviations 
from the isoelectric point. Increasing the attraction among neutral macromolecules even further leads to 
coexisting phases which are approximately composed of neutral macromolecules and solvent in a range of 
pH close to the isoelectric point (Fig.~\ref{fig:phase_diagrams_away_from_pI_both_interactions}(c)). 
Moreover, two discontinuous phase transition points emerge while the two critical points merge and 
vanish. 
In contrast to the previous two cases (a,b) the broadest range in $\bar n$ where phase separation 
occurs is not located at the isoelectric point (c). We must recognise that having such symmetric 
phase diagrams for pH deviations below and above the pI is a consequence of our choice of 
parameters, i.e.\ the charged macromolecules $\mathrm{M^+}$ and $\mathrm{M^-}$ have equal molecular 
volumes and no interactions with the remaining components. Note that the internal free energies 
$w_{\mathrm{M^+}}$ and $w_{\mathrm{M^-}}$ only affect the value of the pI but not the 
symmetry of the phase diagrams around the pI.

We now discuss the effects of pH variations for a system in which charged macromolecules are 
energetically favoured over neutral ones.  We start considering a mixture without neutral-neutral 
interactions (Fig.~\ref{fig:phase_diagrams_away_from_pI_both_interactions}(d)) that exhibits a 
behaviour at the pI resembling a binary mixture of charged macromolecules and solvent 
(Fig.~\ref{fig:topologies_phase_diagrams}(d,f)). For values of pH away from the 
isoelectric point, we observe a monotonic decrease of the macromolecular order parameter, as well 
as a fairly constant charged fraction composition in both phases until they meet at two symmetric 
critical points. After switching on an attractive interaction among neutral macromolecules  a second 
phase separation region appears at larger  values of the total macromolecular volume fraction $\bar n$ 
(Fig.~\ref{fig:phase_diagrams_away_from_pI_both_interactions}(e)). This region is characterised by 
a high density phase mostly composed of neutral macromolecules coexisting with a phase rich in 
charged macromolecules. These two phases meet at two first order transition points (open symbol, 
Fig.~\ref{fig:phase_diagrams_away_from_pI_both_interactions}(e)). The appearance of this region is 
a consequence of having an attraction among neutral macromolecules and charged macromolecules, 
respectively, favouring phase separation dominantly between both while the solvent is of rather 
similar concentration in the coexisting phases. Interestingly, the two regions behave independently 
from each other when increasing the attraction between macromolecules because each region is 
associated to a different solution of chemical equilibrium 
(Fig.~\ref{fig:law_of_mass_action_common_tangent}(a)). While increasing the attraction further,
the two regions merge (Fig.~\ref{fig:phase_diagrams_away_from_pI_both_interactions}(f)).
This merging leads to a broad region of phase separation corresponding to a large difference in the 
fraction of charged macromolecules and solvent as well as  the vanishing of the two critical 
points. The high density phase is made of neutral macromolecules which coexist with a low density 
phase composed of solvent and charged macromolecules. We show the behaviour of $\psi$ along the binodal 
lines in the same phase diagrams in the Supplementary Material.

One typical feature of most  phase diagrams 
(Fig.~\ref{fig:phase_diagrams_away_from_pI_both_interactions}(a,b,d-f)) is that the broader region of 
phase separation exists in the vicinity of the isoelectric point. The region of phase separation shrinks 
when deviating from the pI due to a decrease in interaction energy among charged macromolecules  
(Fig.~\ref{fig:phase_diagrams_away_from_pI_both_interactions}(a,d)) or in the interaction energy among 
both charged macromolecules and neutral 
molecules (Fig.~\ref{fig:phase_diagrams_away_from_pI_both_interactions}(b,e,f)). There is only clear 
exception among these phase diagrams (Fig.~\ref{fig:phase_diagrams_away_from_pI_both_interactions}(c)), 
where the phase separation region slightly increases for pH values away from the isoelectric point. The 
increase is due to the emergence of another stable chemical branch which lowers the free energy by an 
increase in the mixing entropy in the low density phase and increasing the interaction among neutral 
macromolecules in the high density phase. One interesting feature of the phase diagrams, is that when 
the neutral-neutral interactions become dominant, i.e., the phase behavior at the isoelectric point is 
mainly driven by the interaction among neutral macromolecules, we observe the vanishing of the critical 
points, giving rise to phase diagrams which only have first order transitions. The dominant interaction 
thus defines the topology of the phase diagram as a function of pH.

\begin{figure}[]
 \centering
\includegraphics[width=\textwidth]{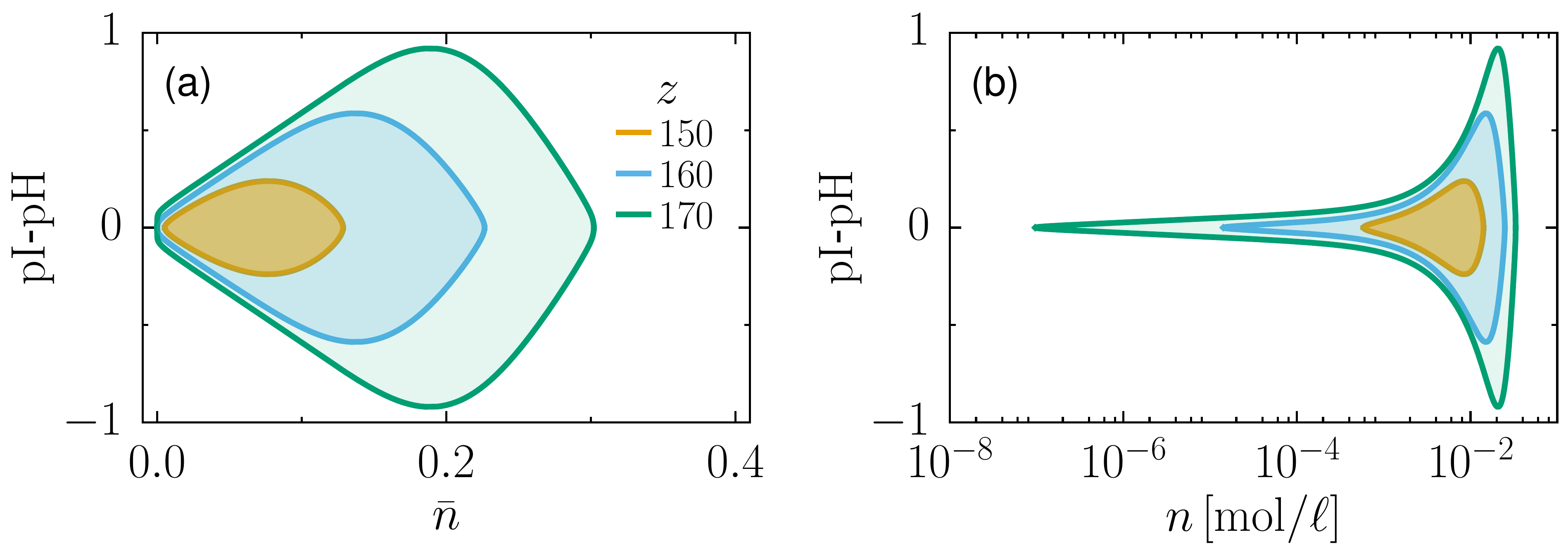}
 \caption{Binodal lines for different choices of the total number of charges on the macromolecules 
$z$ with interactions $\chi_{e}/k_\mathrm{B}T=-\alpha z \epsilon$. The shaded region within the binodals 
is the region where macromolecules undergo a demixing  transition, whereas the region outside is where 
the system remains homogeneously mixed. (a) Phase diagram as a function of the total macromolecule 
volume fraction $\bar n$ and deviations from the isoelectric point $\mathrm{pH-pI}$. (b) Same diagram as 
in (a) but as a function of the total macromolecule molar concentration $n$. Parameters 
$\epsilon=0.002$, $h_\phi/k_\mathrm{B}T=10$ and $\alpha=7.5$, apply to both panels.}
\label{fig:realistic_diagrams}
 \end{figure}

Finally, we study a more realistic scenario where the maximal net charge of the macromolecules $m$ is 
chosen to be $m=50$, which is close to the net maximal charge of some proteins that respond to 
$\mathrm{pH}$ changes and that are found in the so called stress 
granules~\cite{pab1_stress_granules,cellular_fitness_Alberti_2018,footnote_pH1}. For simplicity we only 
consider interactions between oppositely charged macromolecules, which in this case are given by 
$\chi_\mathrm{e}/\epsilon=-\alpha z k_\mathrm{B}T$, where $z$ is the total number of charges of 
the macromolecule and $\alpha$ is a factor describing the contribution of each fixed 
charge of the macromolecules to the interaction (we use the value $\alpha=7.5$ as reported in 
\cite{counterion_release_in_membrane_harries}). We study the system for three different values of the 
total number of charges in the macromolecule $z$ and choosing $\epsilon=0.002$ 
(Fig.~\ref{fig:realistic_diagrams}) motivated by a volume ratio of water molecules and a typical 
protein. For all values of $z$ considered, there exists a broad region of phase separation at the 
isoelectric point (Fig.~\ref{fig:realistic_diagrams}). The coexistence becomes broader for increasing 
values of $z$ due to an increase in the interaction strength. Our minimal 
model predicts that at the isoelectric point, a mixture of 
macromolecules with a large total number of charges, will phase separate over a large concentration 
range. Finally, we also show that reducing $z$ can lead to a drastic reduction of the concentration 
range in which the system undergoes phase separation 
(Fig.~\ref{fig:realistic_diagrams}(b)).

\section{Discussion}\label{sect:discussion}
In this manuscript, we have established a thermodynamic framework to study liquid-liquid phase 
separation in a system where the pH is controlled. We started by introducing chemical reactions 
controlling the charge states of macromolecules which are in turn  determined by the pH value of the 
mixture. Using conservation laws for a system undergoing chemical reactions, we identified the effective 
thermodynamic conjugate variables at chemical equilibrium. We then found the relevant 
thermodynamic variables controlling the pH of the system, namely the chemical potential difference 
$\mu_\mathrm{H_3O^+}-\mu_\mathrm{H_2O}$ and the temperature $T$. That allowed us to 
construct the corresponding thermodynamic potentials for a system with a fixed pH value by means of 
a Legendre transform which makes $\mu_\mathrm{H_3O^+}-\mu_\mathrm{H_2O}$ a natural variable of the 
corresponding free energy. Based on these thermodynamic potentials we showed how the chemical and 
phase equilibrium are controlled by the pH of the system and calculated the corresponding phase diagrams.

We found that phase separation typically occurs around pH values corresponding to the isoelectric 
point pI. Our results could be relevant to cell behavior, where there is a protein spectrum, in 
which many cytosolic proteins have isoelectric points around pH = 5. These results are consistent with 
observations in yeast cells~\cite{liquid_to_solid_Alberti_2016} where many proteins separate from the 
cytosol once the pH is lowered to pH=5. Using typical parameters for cellular proteins 
(Fig.~\ref{fig:realistic_diagrams}) we find that phase separation occurs from concentrations ranging 
from $\mu\mathrm{M}$ corresponding to a typical saturation concentration of phase separating cellular 
proteins~\cite{saha2016polar,cellular_fitness_Alberti_2018} to $\mathrm{m}\mathrm{M}$. Our model is also 
in agreement with phase separation for a concentration of the order of~$1\,\mathrm{mM}$, which is an 
estimation of the total concentration of proteins in yeast~\cite{milo_protein_concentration}. Note that 
phase separation is lost both at low and high concentrations, which can be understood from the analogy 
with the quasi-binary mixture shown in Fig.~\eqref{fig:topologies_phase_diagrams}(a,e). We further 
predict that upon decreasing pH even more, reentrant behaviour leading to a mixed state will be 
observed. However, this behaviour may not always be observable in living systems due to several 
reasons. One of them is that upon decreasing the pH below the isoelectric point, 
proteins might denature and aggregate irreversibly before reaching conditions for phase separation that 
we describe in this manuscript. Our approach considers phase separation controlled by pH at 
thermodynamic equilibrium. Neglecting the effect on phase separation related to the consumption of ATP 
is reasonable since the reported starvation induced phase separation occurs upon ATP depletion 
conditions~\cite{liquid_to_solid_Alberti_2016}. In the future we will consider to extend our approach to 
out of equilibrium situations.

\section*{Acknowledgments}
V.Z. acknowledges financial support from Volkswagen Foundation ``Life?" initiative.
\newpage


\begin{thebibliography}{27}
\bibitem [{foo()}]{footnote_pH1}%
  \BibitemOpen
  \href@noop {} {\ }\bibinfo {note} {The proteins Sup35 and Pab1, which respond
  to pH changes have net charges within this range and the total number of
  charges is approximately as high as $184$ for Sup35.}\BibitemShut {Stop}%
\end{thebibliography}
%

\section*{Supplementary Material}
\renewcommand{\theequation}{S.\arabic{equation}}
\setcounter{equation}{0}
\appendix
\subsection{pH in diluted systems}\label{appendix:pH_definition}

We now show that our definition of pH given in Eq.~\eqref{eq:pH}) is equivalent to the 
most commonly used definition 
$\text{pH}=-\log_{10}(n_{\mathrm{H}_3\mathrm{O}^+}/n_{\mathrm{H}_3\mathrm{O}^+}
^0)$, which only applies for ideal solutions of $\mathrm{H}_3\mathrm{O}^+$ and 
$\mathrm{OH}^-$ in water. In the absence of macromolecules and considering ideal solution 
conditions, the chemical potentials of water $\mu_{\mathrm{H}_2\mathrm{O}}$ and of hydronium ions
$\mu_{\mathrm{H}_3\mathrm{O}^+}$ can be expressed as
\begin{eqnarray}
 \mu_{\mathrm{H}_2\mathrm{O}}&=&v_{0}P+w_{\mathrm{H}_2\mathrm{O}} \quad , 
\label{eq:mu_h20_reg_sol}\\
\mu_{\mathrm{H}_3\mathrm{O}^+}&=&k_\mathrm{B}T\ln(v_{0}n_{\mathrm{H}_3\mathrm{O}^+}
)+v_{\mathrm{H_3O}^+} P+w_ { \mathrm { H } _3\mathrm{O}^+} \quad , 
\label{eq:mu_h3op_reg_sol}
\end{eqnarray}
where we assumed that the contribution of the ions to the volume $V$ is negligible, i.e. taking the 
volume fraction of water $v_0n_\mathrm{H_2O}=1$. The standard chemical potentials 
$\mu_{\mathrm{H}_2\mathrm{O}}^0$ and $\mu_{\mathrm{H}_3\mathrm{O}^+}^0$ are given by the following 
expressions
\begin{eqnarray}
  \mu_{\mathrm{H}_2\mathrm{O}}^0&=&v_{0}P+w_{\mathrm{H}_2\mathrm{O}} \quad , 
\label{eq:mu_h20_standardl}\\
\mu_{\mathrm{H}_3\mathrm{O}^+}^0&=&k_\mathrm{B}T\ln(v_{0}n_{\mathrm{H}_3\mathrm{O}^+}
^ 0
)+v_{\mathrm{H_3O}^+} P+w_ { \mathrm { H } _3\mathrm{O}^+} \quad .
\label{eq:mu_standard}
\end{eqnarray}
Substituting Eqs.~\eqref{eq:mu_h20_reg_sol}-\eqref{eq:mu_standard} in the definition given in 
Eq.~\eqref{eq:pH} we obtain 
\begin{equation}
\mathrm{pH}=-\log_{10}\left(\frac{n_{\mathrm{H}_3\mathrm{O}^+}}{n_{\mathrm{H}_3\mathrm{O}^+}^ 0}
\right) \quad .
\end{equation}
\subsection{Chemical potentials}\label{appendix_exchange_chem_pot}

Here we write the explicit form of the chemical potentials, which are given by
\begin{subequations}\label{eq:chemical_potentials_of_all}
\begin{eqnarray}
\mu_\mathrm{M} &=&  k_\mathrm{B}  T\left (\ln (v\, n_\mathrm{M} 
) + 1\right)+ 2\frac{v}{\epsilon} \chi_{\mathrm{n}} n_\mathrm{M} +w_\mathrm{M}+ v(P-\Sigma) 
\label{eq:mu_rel_m} \quad ,\\
\mu_{\mathrm{M}^+}&= & k_\mathrm{B}  T\left (\ln (v\,  n_\mathrm{M^+})+ 
1\right)+ \frac{v}{\epsilon} 
\chi_{\mathrm{e}} n_\mathrm{M^-} +w_\mathrm{M^+}+ v(P-\Sigma)
\label{eq:mu_rel_mplus} \quad , \\
\mu_{\mathrm{M}^-}&= & k_\mathrm{B}  T\left (\ln (v\, n_\mathrm{M^-}) + 
1\right)+ \frac{v}{\epsilon}  
\chi_{\mathrm{e}} n_\mathrm{M^+} +w_\mathrm{M^-}+ v(P-\Sigma)
\label{eq:mu_rel_mminus} \quad , \\
\mu_{\mathrm{H}_3\mathrm{O}^+} &=& k_{B}T \left(\ln 
\left(v_0 n_{\mathrm{H}_3\mathrm{O}^+}\right) +1 \right) +w_{\mathrm{H}_3\mathrm{O}^+} + 
v_\mathrm{0}(P-\Sigma)
\label{eq:mu_rel_hydronium} \quad , \\
\mu_{\mathrm{OH}^-} &=& k_{B}T \left(\ln 
\left(v_0 n_{\mathrm{OH^-}}\right) +1 \right) +w_{\mathrm{OH^-}} + v_\mathrm{0}(P-\Sigma)
\label{eq:mu_rel_hydroxide} \quad , \\
\mu_{\mathrm{H_2O}} &=& k_{B}T \left(\ln 
\left(v_0 n_{\mathrm{H_2O}}\right) +1 \right) +w_{\mathrm{H_2O}} + v_\mathrm{0}(P-\Sigma)
\label{eq:mu_rel_water} \quad .
\end{eqnarray}
\end{subequations}
\subsection{Choice of conserved quantities}\label{appendix:pH_ensemble}

From the reaction scheme \eqref{eq:reactions} we identify three conserved components in every 
chemical reaction, which 
are
\begin{eqnarray}
N &=& N_{\mathrm{M}^+}+N_{\mathrm{M}^-}+N_\mathrm{M} \label{eq:conserved_number_of_macromolecules} 
\quad , \\
N_H &=& 3N_{\mathrm{H}_3\mathrm{O}^+}+m(2 
N_{\mathrm{M}^+}+N_\mathrm{M})\nonumber 
\\&&+N_{\mathrm{OH}^-}+2N_{\mathrm{H}_2\mathrm{O } } \label { 
eq:conserved_protons} 
\quad , \\
N_s &=& N_{\mathrm{H}_3\mathrm{O}^+} + N_{\mathrm{OH}^-} + N_{\mathrm{H}_2\mathrm{O}} 
\label{eq:conserved_oxygen} \quad 
,
\end{eqnarray}
the three conserved components are:~$N$ the total number of 
macromolecules,~$N_H$ the number of hydrogen atoms 
and~$N_s$ is the number of oxygen atoms. We can combine these equations to show that the partial 
charge involved in the chemical reactions~$N_q=N_H-2N_s-mN$ is also a constant given by 
\begin{equation}
 N_q=N_{\mathrm{H}_3\mathrm{O}^+}+m(N_{\mathrm{M}^+}-N_{\mathrm{M}^-})-N_{\mathrm{OH}^-} \label 
{eq:conserved_charge} \quad .
\end{equation}

\subsection{Concentration of water and its ions at chemical 
equilibrium}\label{appendix:water_comp_chem_eq}

Using Eq.~\eqref{eq:rel_chem_pot_self_ion_water} and 
the chemical potential difference 
$\mu_{\mathrm{H}_{3}\mathrm{O}^+}-\mu_{\mathrm{H}_{2}\mathrm{O}}$, we introduce two more fields 
controlling the relative concentrations of hydronium and hydroxide ions with respect to water 
molecules, these fields obey

\begin{eqnarray}
\log\left(\frac{n_{\mathrm{H}_3\mathrm{O}^+}}{n_{\mathrm{H}_2\mathrm{O}}}\right) &=& 
\frac{h_{\mathrm{H}}}{k_\mathrm{B}T} 
+\frac{h_\psi}{2mk_\mathrm{B}T}\quad , \label{eq:H3O_solvent_field}  \\
\log\left(\frac{n_{\mathrm{OH}^-}}{n_{\mathrm{H}_2\mathrm{O}}}\right) &=&  
\frac{h_{\mathrm{O}}}{k_\mathrm{B}T}
-\frac{h_\psi}{2mk_\mathrm{B}T}\quad , 
\label{eq:OH_solvent_field}
\end{eqnarray}
where the fields $h_{\mathrm{H}}$ and $h_{\mathrm{O}}$ are defined by
\begin{eqnarray}
 h_{\mathrm{H}}&=&\frac{w_{\mathrm{M^+}}-w_{\mathrm{M^-}}}{2m}  
-w_{\mathrm{H}_3\mathrm{O}^+}+w_{\mathrm{H}_2\mathrm{O}} \quad , 
\label{eq:solvent_field_definition_H}  \\ 
h_{\mathrm{O}}&=&\frac{w_{\mathrm{M^-}}-w_{\mathrm{M^+}}}{2m}-w_{\mathrm{OH}^-} 
+ w_{\mathrm{H}_2\mathrm{O}} 
\quad .
\label{eq:solvent_field_definition_O}
\end{eqnarray}
We can go further and use the condition $v_0n_s+vn=1$ and 
Eqs.~\eqref{eq:H3O_solvent_field} and \eqref{eq:OH_solvent_field} to express the concentrations of 
hydronium ions, hydroxide ions and water molecules as a function of the total macromolecule 
density $n$, temperature $T$ and pH (or $h_\psi$), given by
\begin{eqnarray}
n_{\mathrm{H}_2\mathrm{O}}&=&\frac{1-vn}{v_0\left(1+e^{h_\mathrm{H}/k_\mathrm{B}T
+h_\psi/2mk_\mathrm{B}T } +e^ { h_\mathrm{O}/k_\mathrm{B}T-h_\psi/2mk_\mathrm{B}T}\right)} \quad 
,\label{eq:water_density}\\
n_{\mathrm{H}_3\mathrm{O}^+}&=&\frac{(1-vn)\, 
e^{h_\mathrm{H}/k_\mathrm{B}T+h_\psi/2mk_\mathrm{B}T}}{v_0\left(1+e^ 
{h_\mathrm{H}/k_\mathrm{B}T+h_\psi/2mk_\mathrm{B}T} +e^{ 
h_\mathrm{O}/k_\mathrm{B}T-h_\psi/2mk_\mathrm{B}T}\right)} \quad ,\label{eq:hydronium_density}\\
n_{\mathrm{OH}^-}&=&\frac{(1-vn)\, e^{ 
h_\mathrm{O}/k_\mathrm{B}T-h_\psi/2mk_\mathrm{B}T}}{v_0\left(1+e^
{h_\mathrm{H}/k_\mathrm{B}T+h_\psi/2mk_\mathrm{B}T} +e^{ 
h_\mathrm{O}/k_\mathrm{B}T-h_\psi/2mk_\mathrm{B}T}\right)} \quad 
.\label{eq:hydroxide_density}
\end{eqnarray}

\subsection{Critical points at the isoelectric point}\label{appendix:critical_behavior}
In this section we calculate some limiting critical values at the 
isoelectric point, where $\psi=0$ and $h_\psi=0$.

\subsubsection{Effective binary critical point}\label{appendix:binary_regime}
In order to calculate the critical points describing the effective binary mixture, we first 
discuss the limits of $h_\phi$ for $\phi \to 1/2$ and $\phi \to 0$, which correspond to situations 
where macromolecules are only charged or only neutral respectively.

If we consider the limit of $h_\phi$ with $\phi \to 1/2$, 
\begin{align}
\nonumber
\lim_{\phi\to \frac{1}{2}} h_{\phi}=& \lim_{\phi\to \frac{1}{2}} \left[2k_\mathrm{B}T\ln 
\frac{\phi}{1-2\phi}  + \chi \bar n (\phi -\lambda)/\epsilon \right] \quad ,\\
\nonumber
\lim_{\phi\to \frac{1}{2}} h_{\phi}=& -2\ln 2 -2k_\mathrm{B}T \lim_{\phi\to \frac{1}{2}} 
\ln(1-2\phi) +\frac{\chi \bar n}{\epsilon} \left(\frac{1}{2} -\lambda \right) \quad ,\\
\nonumber
\lim_{\phi\to \frac{1}{2}} h_{\phi}=& \infty \quad ,
\end{align}
this shows on one hand that large positive values of $h_\phi$ are obtained for values of $\phi$ 
approaching $1/2$, on the other hand, large negative values of $h_\phi$ are obtained for $\phi$ 
approaching $0$. 
\begin{equation}
 \nonumber \lim_{\phi \to 0} h_\phi=-\infty \quad .
\end{equation}

We can study both cases by direct substitution in the free energy density 
\eqref{eq:free_energy_density_f}. We 
illustrate the case for $\phi=1/2$, in this case the free energy density reads:
\begin{equation}
v \bar f = k_{B}T \left( \bar n \ln \left( \frac{\bar n}{2}\right) + 
\frac{(1-\bar n)}{{\epsilon}}\ln (1-\bar n)\right) + \frac{\chi_{\mathrm{e}} \bar n^2}{4\epsilon} + 
\mathcal{O}(\bar n) \, , 
\label{eq:reduced_free_energy}
\end{equation}
the linear terms~$\mathcal{O}(\bar n)$ do not affect the stability of the system, therefore we are 
safe to ignore them in our calculation. From the free energy evaluated at $\phi=1/2$ 
\eqref{eq:reduced_free_energy}, we calculate the 
chemical potential up to a constant
\begin{equation}
\label{eq:mu_effective}
 v \frac{d\bar f}{d\bar n}=k_{B}T\left[ \ln \left( \frac{\bar n}{2}\right)+1 
-\frac{\ln(1-\bar 
n) +1}{\epsilon}+\frac{\chi_\mathrm{e}\bar n}{2\epsilon}\right]
\end{equation}
using Eq. \eqref{eq:mu_effective} and conditions~$d^2 f/d\bar n^2=0$ and~$
d^{3}f/d\bar n^{3}=0$ we find 
\begin{eqnarray}
\bar n_c^{b}&=&\frac{\sqrt{\epsilon}}{1+\sqrt{\epsilon}}  
\quad ,\label{eq:n_e_b} \\
k_\mathrm{B}T_c &=& -\frac{\chi_\mathrm{e}}{2(1+\sqrt{\epsilon})^2} \quad  .
\end{eqnarray}
Following the same procedure for a system with~$\phi = 0$ gives
\begin{eqnarray}
\bar n_c^{b}&=& \frac{\sqrt{\epsilon}}{1+\sqrt{\epsilon}} \quad , \\ 
k_\mathrm{B}T_c &=& -2 \frac{\chi_\mathrm{n}}{(1+\sqrt{\epsilon})^2} \quad  . 
\end{eqnarray}

\subsubsection{Critical point at $\bar n =1$}
Here, we calculate the critical value which emerges at~$\bar n=1$ using the free 
energy density \eqref{eq:free_energy_density_f}, which at the isoelectric point is
\begin{eqnarray}
 v\bar f&=&k_{\mathrm{B}}T\left(2\phi \ln \phi + (1-2\phi)\ln(1-2\phi)\right)+\chi 
\left(\phi^2-2\lambda\phi+\lambda/2 \right)/\epsilon -h_\phi \phi + w_\mathrm{M} \quad .
\end{eqnarray}
We first differentiate $\bar f$ with respect to $\phi$ which gives
\begin{equation}
 v \frac{\partial {\bar f}} {\partial {\phi}}=k_\mathrm{B} T\left(2\ln\phi - 
2\ln(1-2\phi)\right) +2\chi\left(\phi-\lambda\right)/\epsilon-h_\phi  
\label{eq:chem_pot_at_n_1} \quad .
\end{equation}
The conditions for finding a critical point in this case are~$\partial^2 \bar f/\partial 
\phi^2=0$ 
and~$\partial^3 \bar f/\partial \phi^3=0$, these conditions read
\begin{eqnarray}
 k_\mathrm{B}T_c \left(\frac{2}{\phi_c}+\frac{4}{1-2\phi_c}\right)+2\chi/\epsilon=0 
\label{eq:crit_point_n_1_spin}  \quad , \\
 k_\mathrm{B}T_c\left(-\frac{2}{\phi_c^2}+\frac{8}{(1-2\phi_c)^2}\right)=0 
\label{eq:crit_point_n_1_spin_derivative} 
\quad.
\end{eqnarray}
Solving Eqs. \eqref{eq:crit_point_n_1_spin} and \eqref{eq:crit_point_n_1_spin_derivative} we find
\begin{eqnarray}
 \phi_c &=& \frac{1}{4} \label{eq:app_phi_crit_n_1} \quad , \\
 k_\mathrm{B}T_c &=& -\frac{\chi}{8\epsilon} \quad 
,\label{eq:crit_interactions_n_1} \\
h_{\phi,c}&=&\chi \left(\frac{\ln 2 + 2 - 8\lambda}{4\epsilon}\right)
\label{eq:app_crit_mol_field_n_1} \quad .
\end{eqnarray}
\subsection{Phase diagrams as a function of pH}\label{appendix:phase_diagrams_psi_pH}
Here we show the phase diagrams from  Fig.~\ref{fig:phase_diagrams_away_from_pI_both_interactions} 
showing the dependence in $\psi$.
\begin{figure}
 \centering
\includegraphics[width=\textwidth]{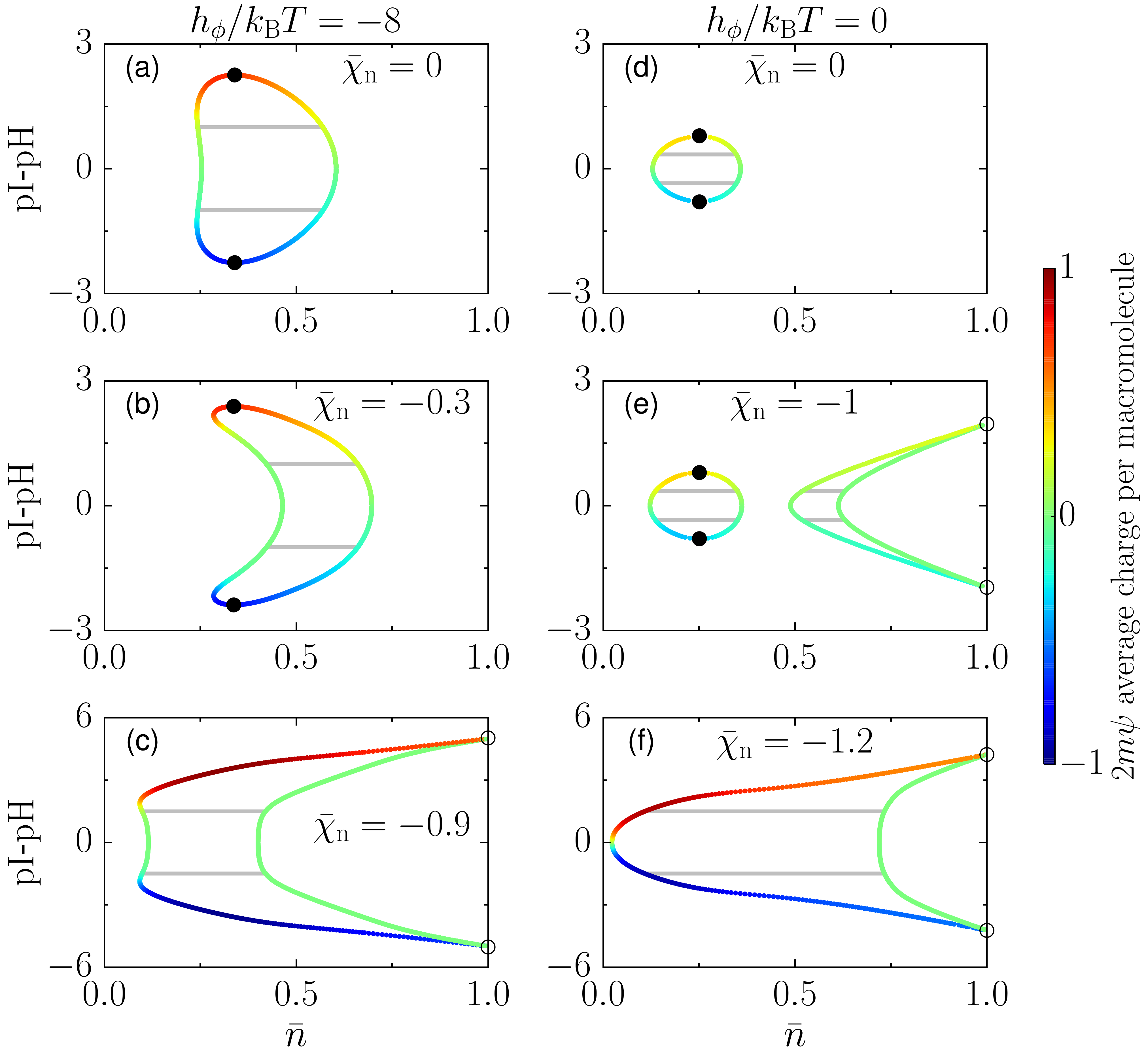}
  \caption{Phase behaviour as a function of pH, the colorbar indicates the average charge per 
macromolecule $2m\psi=m(n_\mathrm{M^+}-n_\mathrm{M^-})/n$. Parameters $\epsilon=0.1$ 
and $\chi_\mathrm{e}/k_\mathrm{B}T=-3.5$, apply to all panels.} 
\label{fig:phase_diagrams_away_from_pI_psi_both_interactions}
 \end{figure}
\end{document}